\documentclass[aps,floatfix,twocolumn,showpacs,superscriptaddress]{revtex4}
\usepackage{graphicx}
\usepackage{wasysym}
\usepackage{amssymb}
\usepackage{bm}
\begin{document}
%
\title{Polyelectrolyte-surfactant complex:\\
phases of self-assembled structures}
\author{C. von Ferber}
\email{ferber@physik.uni-freiburg.de}
\affiliation{Theoretical Polymer Physics, 
Freiburg University, 79104 Freiburg, Germany}
\author{H. L\"owen} 
\affiliation{Institut f\"ur Theoretische Physik II, 
 Heinrich-Heine-Universit\"at D\"usseldorf, 40225 D\"usseldorf,
 Germany}
\begin{abstract} 
We study the structure of complexes formed between ionic surfactants (SF)
and a single oppositely charged polyelectrolyte (PE) chain. For our
computer simulation we use the ``primitive'' electrolyte model: while
the polyelectrolyte is modeled by a tethered chain of charged hard
sphere beads, the surfactant molecules consist of a single charged
head bead tethered to a tail of tethered hard spheres.  A hydrophobic
attraction between the tail beads is introduced by assuming a
Lennard-Jones potential outside the hard-sphere diameter. As a function
of the strengths of both the electrostatic and the hydrophobic
interactions, we find the following scenario: switching on and
increasing the electrostatic forces first leads to a stretching of the
PE and then by condensation of SF to the formation of a complex.  For
vanishing hydrophobic forces this complex has the architecture of a
molecular bottle-brush cylindrically centered around the stretched PE
molecule. Upon increasing the hydrophobic attraction between the SF
tails, a transition occurs inverting this structure to a spherical
micelle with a neutral core of SF tails and a charged corona of SF
heads with the PE molecule wrapped around. 
At intermediate hydrophobicity there is a competition between the two
structures indicated by a non-monotonic dependence of the shape
as function of the Coulomb strength, favoring the cylindrical
shape for weak and the spherical micellar complex for strong
interaction. 
\end{abstract}
\date{\today}
\maketitle

\section{Introduction}\label{I}
Polyelectrolyte-surfactant mixtures have proven to provide the basis
for new materials with extraordinary properties that make them
interesting for a wide range of applications
\cite{Kwak-1998,Antonietti-1997}.  In particular the Coulomb
attraction between polyelectrolyte chains and oppositely charged ionic
surfactant molecules in solution leads to aggregation and complex
formation.  The resulting complex differs in its conformational,
structural and dynamical features from that of solution containing only
one of its constituents, the pure polyelectrolyte or
the pure surfactant (for reviews see
\cite{Hansson-1996,Kwak-1998}).
Much experimental work has been done on these systems and a rich
variety of different complex structures was revealed as a function of
chemical nature of the polyelectrolyte, surfactant and solvent
molecules
\cite{Sokolov-1998,Kosmella_1998,Claesson_1998,Tsianou-1999,
Thunemann-2000a,Dias-2000,Babak-2000,Liao-2001,Guan-2001,
Taylor-2002,Hansson-2002,Guillot-2003}.
A theory that could give 
a systematic microscopic characterization and prediction of the different
complexes on the other hand does not exist so far.

Already for complexes formed by a {\it single} polyelectrolyte chain
with ionic and hydrophobic surfactants at low concentration a
theoretic treatment that takes the long-ranged Coulomb interaction and
the hydrophobic interactions fully into account is still to be
developed. The case without hydrophobicity has recently 
been simulated by the present authors \cite{Ferber-2003}.
For these single polyelectrolyte complexes one expects
structures either of bottle-brush shape for low hydrophobicity or, for
strong hydrophobicity, possibly spherical micelles that aggregate
together with the polyelectrolyte chain
\cite{Antonietti-1997,Wallin-1998}.  Theories so far in general make
assumptions about the symmetry or the structure of the complex
implying a complete adsorption of surfactant onto the polyelectrolytes
for rigid \cite{Shira1,Shira2} and flexible chains, see e.g.\
\cite{Sear} and cannot predict the complex structure or its symmetry
\cite{Wallin-1998} as far as it enters the theory as an input.  
For the special case of a system of rigid polyelectrolytes that form
complexes via thermodynamic counterion condensation including the
effect of added surfactant a theory has
recently been developed by Kuhn, Levin and coworkers
\cite{Kuhn-1998,Kuhn-2000,Silva-2001}.
The collapse and partial collapse of semiflexible and flexible
polymers in solutions with surfactants has been treated by a meanfield
association theory by Diamant and Andelman
\cite{Diamant-2000a,Diamant-2000b}. However, this theory does not
include explicitly the long range of the Coulomb interactions.  A self
consistent field theory with a spherically symmetric setup
\cite{Wallin-1998} on the other hand predicts the formation of spherically 
symmetric micelles with a neutral hydrophobic core and a surface layer
where the charged surfactant heads and the oppositely charged
polyelectrolyte are located. Based on this observation it has been
proposed to simplify the model by replacing the ionic surfactant by
large spherical counterions not treating the surfactant tails
explicitly \cite{Wallin-1996}. 
For a system of a polyelectrolyte in the presence of small counterions 
computer simulations
\cite{Kremer1,Winkler,Kremer2} and theories
\cite{Brilliant,Kuhn-2002} find a swelling and stretching of the
polyelectrolyte and a subsequent collapse when increasing 
the Coulomb interaction. In this scenario a non-monotonic 
dependence of the of the polyelectrolyte 
\newpage
\begin{widetext}
\begin{figure}[t]
\includegraphics[width=160mm]{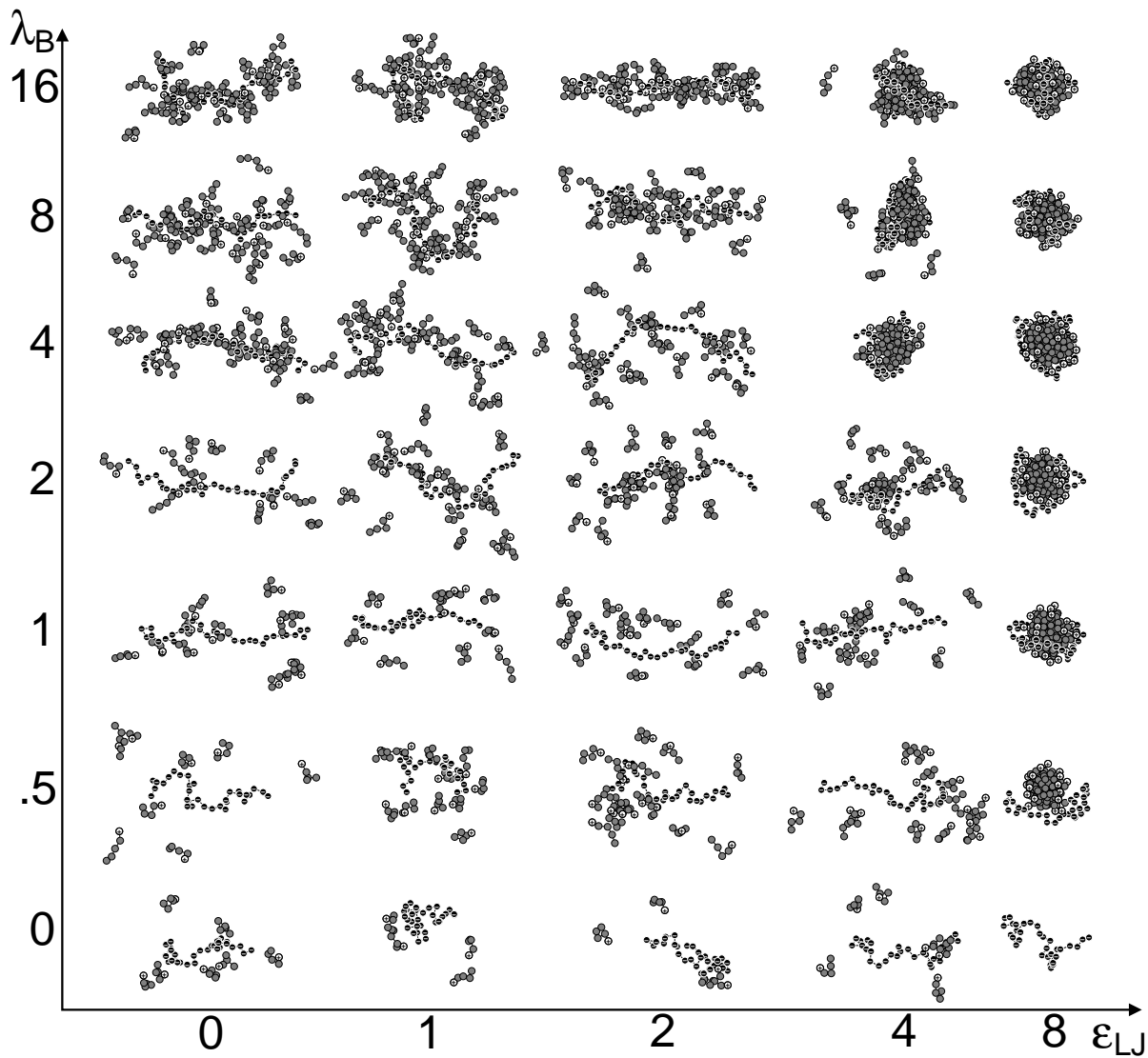}\\
\caption{\label{phasesketch} Snapshots of the complex configurations
showing the dependence on the two interactions: Coulombic ($\lambda_B$) and
hydrophobic ($\varepsilon_{LJ}$)}
\end{figure}
\end{widetext}
conformation as a function
of the Coulomb strength occurs:
for weak Coulomb strength the PE stretches while
in the strong Coulomb regime counterion condensation and a collapse to
a coil conformation occurs.
 For large counterions representing micelles the polyelectrolyte may be wrapped around the charged sphere exhibiting different
conformations depending on the chain flexibility 
\cite{Wallin-1996,Gurovich-1999,Mateescu-1999,Netz-1999,Park-1999,Kunze-2000,Nguyen-2000,Welch-2000,Schiessel-2001,Jonsson-2001,Jonsson-2001a,Akinchina-2002,Chodanowski-2001,Chodanowski-2001a,Brynda-2002,Keren-2002}.

The bottle brush structure that is expected for the condensation of
charged ionic surfactants onto a polyelectrolyte chain may be compared
with corresponding neutral molecular bottle brushes in case the
polarization effects in the complex are neglected. Such neutral bottle
brush or comb polymers are built either by end linking the side
chains to the backbone chain or by attaching them via strong hydrogen bonds
\cite{tenBrinke-1997,Ruokolainen-1996}. Such configurations may be
expected to become more rigid for higher densities of the side chains.
A quantitative description, however, is still under debate since
different simple scaling arguments
\cite{Birshtein-1987,Fredrickson-1993,Rouault-1996} appear to predict
different quantitative behavior for the sizes of the main and side
chains depending on the model used and the limit that is considered
\cite{Khalatur-2000}.

In the present paper we explore by computer simulation the phases of
complex formation of a single polyelectrolyte chain with ionic
surfactants in the phase space parameterized by the strengths of the
Coulombic interactions and the hydrophobic attraction between the
surfactant tails.  As in our previous study \cite{Ferber-2003} our
simulation relies on the so-called ``primitive'' model
\cite{Hansen-2000} that treats the microscopic charges of the
polyelectrolyte and the surfactant heads explicitly. This is in
contrast to simulations on polymer-surfactant complexation which
neglect the long-range nature of the Coulomb interactions
\cite{Groot-2000}.  The ``primitive'' model has recently also been applied
to simulate the complexation of polyelectrolyte chains with charged
spheres representing spherical micelles \cite{Rescic-2000} and has
proven to be reliable in different contexts of polyelectrolyte
conformations \cite{Kremer1, Winkler,Kremer2,Wallin-1997,
Winkler-2002,Jusufi-2002,Messina-2002a}.

To treat surfactants with neutral or hydrophobic tails we 
model the surfactant by a tethered hard sphere chain
with a charged head and a Lennard-Jones like effective attraction
between the hydrophobic tail monomers allowing for the formation of
micelles.  As a result, we observe different phases of complexation
depending on the strengths of the two interactions involved: the
Coulomb and the hydrophobic interaction.  An overview of these phases
is given in Fig. \ref{phasesketch} in terms of simulation snapshots.
For vanishing hydrophobicity (leftmost column of
Fig. \ref{phasesketch}) the mean square end-end distance $R^2$ of the
polyelectrolyte is non-monotonic as function of the Coulomb
interaction. However, in contrast to the standard situation where a
collapse of the polyelectrolyte due to counterion condensation occurs
\cite{Kremer1,Winkler,Kremer2}, such a collapse
is blocked by the steric interaction of the surfactant tails.  As we
have shown in Ref. \cite{Ferber-2003}, for even longer surfactant
tails the collapse is more sufficiently blocked, while the addition of
salt weakens the complex as it replaces the surfactant in the complex.
Increasing the hydrophobicity for vanishing Coulomb interaction on the
other hand (bottom row of Fig.\ref{phasesketch}) leads to the
aggregation of micellar clusters of surfactant molecules that are
independent from the polyelectrolyte chain. Introducing a finite
Coulomb strength in this latter situation favors the condensation of
the surfactant at the polyelectrolyte. Thirdly we may start with a
bottle brush conformation of the complex at high Coulomb but vanishing
hydrophobic interactions (top left of
Fig.\ref{phasesketch}). Gradually increasing the hydrophobicity (along
the top row of Fig.\ref{phasesketch}) the tails start to aggregate and
and a transition occurs from the cylindrical structure with the tails
on the outside to an inverted spherical structure with a core that is
constituted by the surfactant tails.  In our present resolution this
transition appears to be sharp.

The paper is organized as follows: in section II, we describe our
model in detail and we present details of the simulation procedure in
section III.  Results are given in section IV and we conclude in
section V.


\section{The Model}\label{II}
In our simulation we model all molecules as chains of tethered hard
spheres (beads) that represent Kuhnian segments of the
molecule. Microions are modeled by hard sphere beads. Charges on these
molecules and ions are represented by placing a point charges at the
center of any charged bead. In this primitive model the solvent is not
treated explicitly and only enters via its dielectric constant
$\epsilon_r$ neglecting the discreteness of the solvent molecules.
The microscopic interactions between the particles are given by the
hard sphere excluded volume, the long-ranged Coulomb forces, and a
short range hydrophobic interaction.  We reduce the model parameter
space by fixing the hard sphere diameter of all beads in the system to
the same value $\sigma$.  Furthermore, all charged beads have the same
charge $\pm q e$ with $e$ denoting the elementary charge.  While this
``minimal'' model explicitly handles the counter ions and may be used
to describe strongly charged, flexible
polyelectrolytes \cite{Kremer1}, analytical calculations usually
neglect the individual character of the counter ions
\cite{Odijk-1979,Fixman-1978,deGennes-1976,Barrat-1993}.

\begin{figure}[t]
\includegraphics[width=82mm]{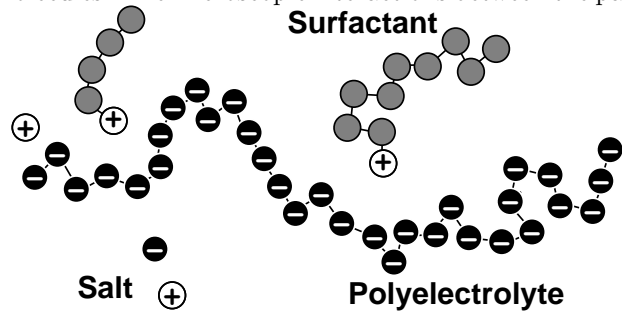}\hspace{3em}
\caption{\label{fig1}
Model of the polyelectrolyte - surfactant system with
salt ions. Charges are indicated by "+" and "-" signs.\\[-10mm]
}
\end{figure}

As mentioned above, all molecules are modeled by chains of freely
jointed, tethered hard sphere beads. Subsequent beads in the same
chain are tethered.  The extension of all bonds is limited by the
tether to a fixed value $b$.  For vanishing Coulomb and hydrophobic
interactions the model is purely entropic (i.e.\
temperature-independent).  Our specific parameters are such that the
polyelectrolyte is represented by a chain of $N$ charged beads each
with charge $-qe$.  The surfactant molecules on the other hand are
chains of $n$ tethered beads with one charged head bead of charge
$+qe$ and $n-1$ neutral tail beads that may interact due to hydrophobicity.
The full model may also contain salt which has been treated
in Ref.\cite{Ferber-2003}. The salt ions are then represented
by single charged beads. 
The constituents of the model of polyelectrolytes, surfactant
molecules and microions is schematically shown in Figure \ref{fig1}.

In our model,
the effective interaction due to hydrophobicity of the surfactant
tails is given by a Lennard-Jones like potential between the tail
monomers.  This approach has also been used in the context of
copolymer micellization
\cite{Connolly-2003,Timoshenko-2000}.
The explicit interaction between any two beads $i,j$ of the model
at a distance $r$ between their centers is given by a potential
$V_{ij}(r)$. To denote chain connectivity we use 
 an adjacency matrix $T_{ij}$. For
beads $i,j$ that are tethered $T_{ij}=1$ otherwise $T_{ij}=0$.  With
$q_i \in \{\pm 1,0\}$ denoting the charge
and $u_i \in \{1,0\}$ the hydrophobicity number of bead $i$,
the potential $V_{ij}(r)$ is given by
\begin{widetext}
\vspace*{-5mm}
\begin{equation} \label{1}
\frac{1}{k_{\rm B}T}V_{ij}(r)=
 q_iq_j\lambda_{\rm B}\frac{\sigma}{ r}
 +u_iu_j\varepsilon_{LJ}
(\left(\frac{\sigma}{ r}\right)^{12}
-\left(\frac{\sigma}{ r}\right)^{6})
 +V_{\rm H}(\sigma-r) + T_{ij}V_{\rm H}(r-b).
\end{equation} 
\end{widetext}
Here, the dimensionless Coulomb strength $\lambda_{\rm B}$ is the Bjerrum
length measured in bead diameters $\sigma$.
The Bjerrum length
$\sigma\lambda_{\rm B}={q^{2}e^2}/{\epsilon_r k_{\rm B}T}$ sets the length scale
where the Coulomb pair interaction is comparable to the thermal
energy $k_BT$. The dimensionless hydrophobicity parameter $\varepsilon_{LJ}$ 
controls the depth of the Lennard-Jones potential.
Furthermore, the hard core and tether interactions are
represented by the one dimensional hard wall potential
\begin{equation}\label{2}
V_{\rm H}(x)=\left\{
\begin{array}{rl} 
0 & \mbox{ if } x<0 \\
\infty & \mbox { else }
\end{array}\right.\ .
\end{equation}
We note that in our notation the product $T_{ij}V_{\rm H}(r-b)$ is 
defined to be zero if $T_{ij}$ is zero.

The key parameters of our model are: the number $N$ of charged
monomers of the polyelectrolyte chain, the number $n-1$ of neutral
beads of the surfactant tails, the parameters $\lambda_B$ and
$\varepsilon_{LJ}$ which measure the strengths of the Coulomb and
hydrophobic interactions and finally the relative salt concentration
$c_s$.  In the following we shall explore the parameter space in
particular in the two variables $\lambda_B$ and $\varepsilon_{LJ}$
while keeping fixed the particle hard-core diameters, the microion
charges $q_i$, the monomer number $N=32$ and the maximal tether length
$b=1.5\sigma$ and in general also the salt concentration $c_s=0$ and
the surfactant length $n=5$.  In our previous work \cite{Ferber-2003}
we also investigated the influence of surfactant tail variation as well as
of finite salt concentration. A brief discussion of the
corresponding results is also included in the following.
\section{Simulation Technique}\label{III}

We simulate by standard Monte Carlo (MC) methods a system of a single
polyelectrolyte chain of $N$ charged beads at finite concentration
with polyelectrolyte charge density $\rho=N/L^3=0.001\sigma^{-3}$
which is kept fixed throughout all simulations.  The finite
concentration is taken into account by periodic boundary conditions of
our finite cubic simulation box of length $L$ and replicated images of
the charges.  Summations over the interactions with the periodic
images are calculated by the Lekner sum method \cite{Lekner-1991}. The
Metropolis rates of the MC moves are determined by the interactions
given by Eqn.\ (\ref{1}).

The relaxation of the system, especially in the fully assembled bottle
brush configurations is very slow. This is also known from simulations
of conventional bottle brush molecules \cite{Khalatur-2000}. To
ensure sufficient relaxation we performed 
$\sim 5\cdot 10^6$ 
attempted MC moves per particle at each state point.
 The acceptance ratio is roughly $0.8$
in situations with an open structure. In the micellar cases however,
caging occurs inside the densely packed core reducing the acceptance
rate down to $0.3$.

In all simulations the global charge of our system vanishes.
We performed simulations for a
system of a single polyelectrolyte chain together with $N=32$
oppositely charged surfactant molecules.  The case $n=1$ of no tail
beads at all serves as a natural reference for pure counter ions.  Such
a situation was already studied in Ref.\ \cite{Kremer1} 
and implemented in the frames of the present setup in Ref. 
\cite{Ferber-2003} for
different Bjerrum lengths. Note, however, that chain
connectivity was modeled differently in Ref.\ \cite{Kremer1} via
a finite extension potential. 


\section{Results}\label{IV}
\subsection{Zero hydrophobicity of the surfactant}
In the case of vanishing hydrophobic interactions $\varepsilon_{LJ}=0$
the surfactant tails interact only due their steric repulsion. This
special case was treated by the present authors in \cite{Ferber-2003}
in detail, so we only recall some of the basic results.  We first
discuss the case of no added salt.  As a reference situation serves
the a surfactant without tail, i.e. $n=1$.  The averaged square
end-to-end distance $R^2$ of the polyelectrolyte chain is defined via
\begin{equation}\label{3}
R^2 = < {({\vec R}_1 - {\vec R}_N)^2} >
\end{equation}
where $<...>$ denotes a statistical average and ${\vec R}_1$ and
${\vec R}_N$ are the actual positions of the two end-monomers of the
polyelectrolyte.
\begin{figure}
\includegraphics[width=80mm]{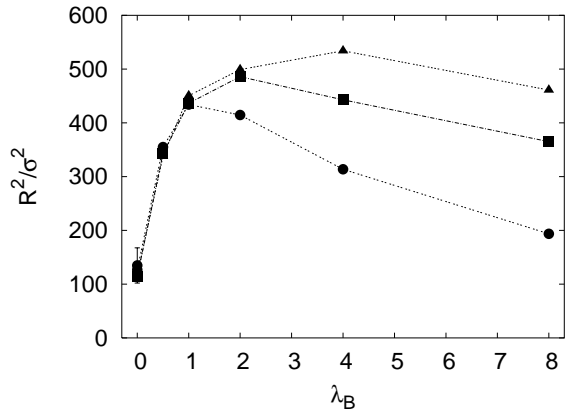}\\
\caption{\label{fig2}
The mean square end-to-end distance $R^2$ as a
function of the Bjerrum length $\lambda_B$ for $n=1 ({\displaystyle \bullet})$,
 $5 (\square)$ and $10(\blacktriangle) $ (from below) monomer
surfactants for vanishing hydrophobicity parameter $\varepsilon_{LJ}=0$. 
The statistical uncertainty is indicated by an error bar.
}
\end{figure}

$R^2$ is plotted as a function of Coulomb
strength $\lambda_B$ in Fig. \ref{fig2}. Here,
the collapse scenario found in other simulations 
\cite{Kremer1,Winkler,Kremer2} and analytical treatments
\cite{Schiessel-1998} is confirmed.
This scenario for simple counterions is the following:
At zero Bjerrum length $\lambda_B=0$ the interaction is purely 
excluded volume and the chain attains a polymer coil configuration.
Increasing $\lambda_B$ the chain stretches due to the repulsion
of its charged beads. For low $\lambda_B$ the counterion screening 
is weak and stretching continues until $\sigma\lambda_B$ is of the 
order of the counterion diameter, i.e. $\lambda_B\simeq 1$.
For larger $\lambda_B>1$ the condensation of counterions 
induces a shrinking of the PE extension which for high $\lambda_B$
may even be become smaller than the neutral coil; an effect not
yet seen in the range for $\lambda_B$ that we simulated.

Substituting in the former system the counterions by charged surfactants
with neutral tails the behavior as function of the Coulomb strength 
$\lambda_B$ changes in the following way: The PE chain stretches for
increasing $\lambda_B$ with the maximal extension attained for
higher values of $\lambda_B$ corresponding the the larger effective
diameter of the charged surfactant acting as a counterion. 
When counterion condensation occurs at higher $\lambda_B$ the surfactant
and the PE chain form a complex in which the surfactants aggregate
with their heads near the PE chain and their tails pointing away
from the PE chain in a manner resembling the structure of a 
molecular bottle brush.  
The subsequent
collapse of the PE chain that is observed in the previous scenario
now is effectively blocked due to the excluded volume interaction
of the surfactant tails. 
The internal structure of this complex is discussed below.  For
the special case of vanishing hydrophobicity  the
influence of added salt on the end-to-end distance $R$
was investigated in Ref.\cite{Ferber-2003}. 
\begin{figure}
\includegraphics[width=80mm]{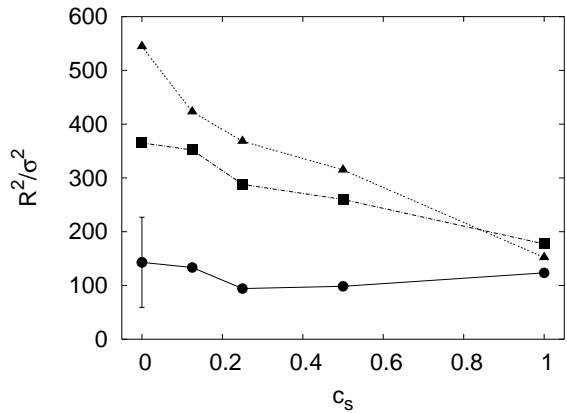}\\
\caption{\label{fig6}
The mean square end-to-end distance $R^2$ for Bjerrum
length $\lambda_B=8\sigma$ as a function of the relative salt
concentration $c_s$ for (from below) $n=1 ({\displaystyle \bullet})$,
 $5 (\square)$ and $10(\blacktriangle) $  monomer surfactants.
The statistical uncertainty is indicated by an error bar.
}
\end{figure}
Adding $N_s$ additional $1:1$ salt pairs to our system
we define the relative concentration as $c_s=N_s/N$ where $N$ is the
number of charges on the polyelectrolyte. The results are shown
in Fig. \ref{fig6}.  The surfactant molecules in the complex are
replaced in favor of salt ions. This is driven by entropy of mixing
and corresponds to a chemical equilibrium. The replacement is
therefore increasing with increasing salt concentration.  This
destroys the structure of the complex as seen from the reduction in
the end-to-end distance of the polyelectrolyte.

In all our simulations we observe that
the contour length $L$ of the
polyelectrolyte given by the sum of the bond lengths
 is very stable under all changes of the
parameters.  Varying the Bjerrum length in
the range $\lambda_B=0\ldots 8$ we find a corresponding increase from
$L=40$ to $L=42$, 
a change that is within the statistical uncertainty.  For this
reason also the persistence length $L_p$ does not give additional
information. For a worm-like chain namely it is related to the end-end
distance $R$ and the contour length $L$ by
\cite{Doi-1986}
\begin{equation}
 R^2 = 2 L L_p-2L_p^2(1-\exp(-L/L_p))
\end{equation}
a relation also confirmed by simulations \cite{Kremer1}.
\subsection{Clusters and micelles}
The results described in this and the following sections section all
refer to a system with a single polyelectrolyte chain of 32 charged
beads and 32 oppositely charged 5-bead surfactant molecules with one
charged head bead and four neutral possibly hydrophobic tail beads.
For finite values of the Coulombic and hydrophobic interactions we
observe in general one or more aggregates or clusters of molecules. To
analyze these structures in detail we define the clusters by assuming
that any two beads belong to the same cluster if their distance is
less than $2\sigma$. Among these clusters we identify the complex as
the one that contains the polyelectrolyte chain.  
\begin{figure}
\includegraphics[width=80mm]{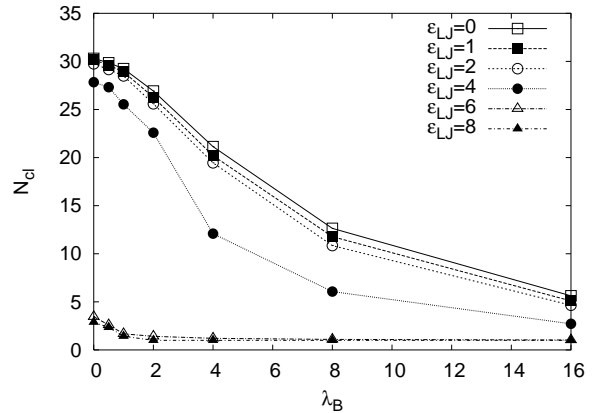}\\
\caption{\label{shapeUM0} The number of clusters $N_{cl}$ in the system 
as function of the Coulomb strength $\lambda_B$ for different values
of the hydrophobicity parameter 
$\varepsilon_{LJ}=0 (\square), 1 (\blacksquare), 2 ({\displaystyle \circ}),
4 ({\displaystyle \bullet}), 6 (\vartriangle), 8 (\blacktriangle)$ from above.}
\end{figure}
We monitor the
number of clusters and the sizes of the complex and of the largest
aggregates that are detached from the polyelectrolyte. These latter
aggregates we identify with isolated micelles. The number of
identified clusters is shown in Fig. \ref{shapeUM0} as a function of the
Coulomb interaction parameter $\lambda_B$ for different values of the
hydrophobicity strength $\varepsilon_{LJ}$: For vanishing interactions
clustering occurs only by chance and almost each cluster observed
represents one of the 33 single molecules in the system.  For
increasing interactions the molecules condense to a few larger
clusters with only one or two remaining in the strong interaction
case.  When the Coulomb interaction vanishes, i.e. $\lambda_B=0$, the
polyelectrolyte will always be detached from the micelles and thus a
minimal number of two clusters remain even for strong hydrophobic
attraction.  For any finite Coulomb interaction and high
hydrophobicity on the other hand all molecules condense to a single
complex in the present setup.  
\begin{figure}
\includegraphics[width=80mm]{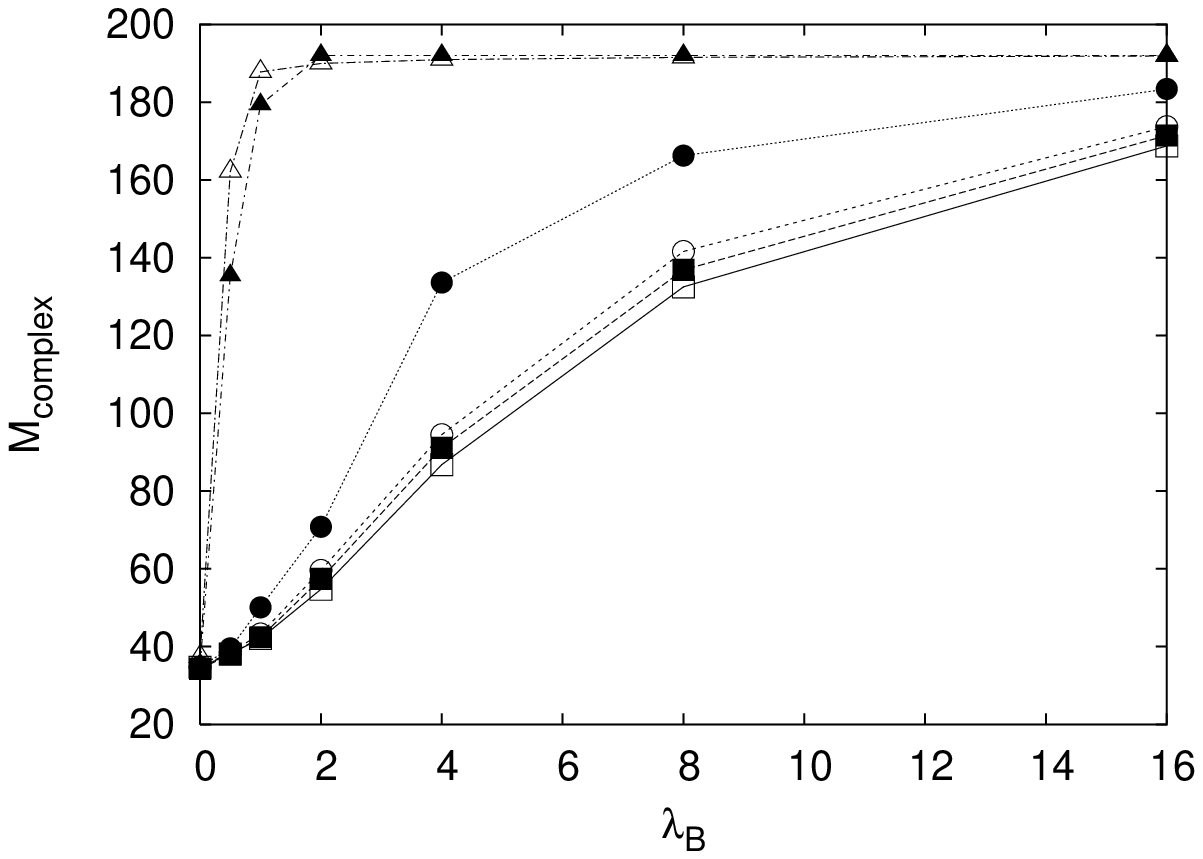}\\
\includegraphics[width=80mm]{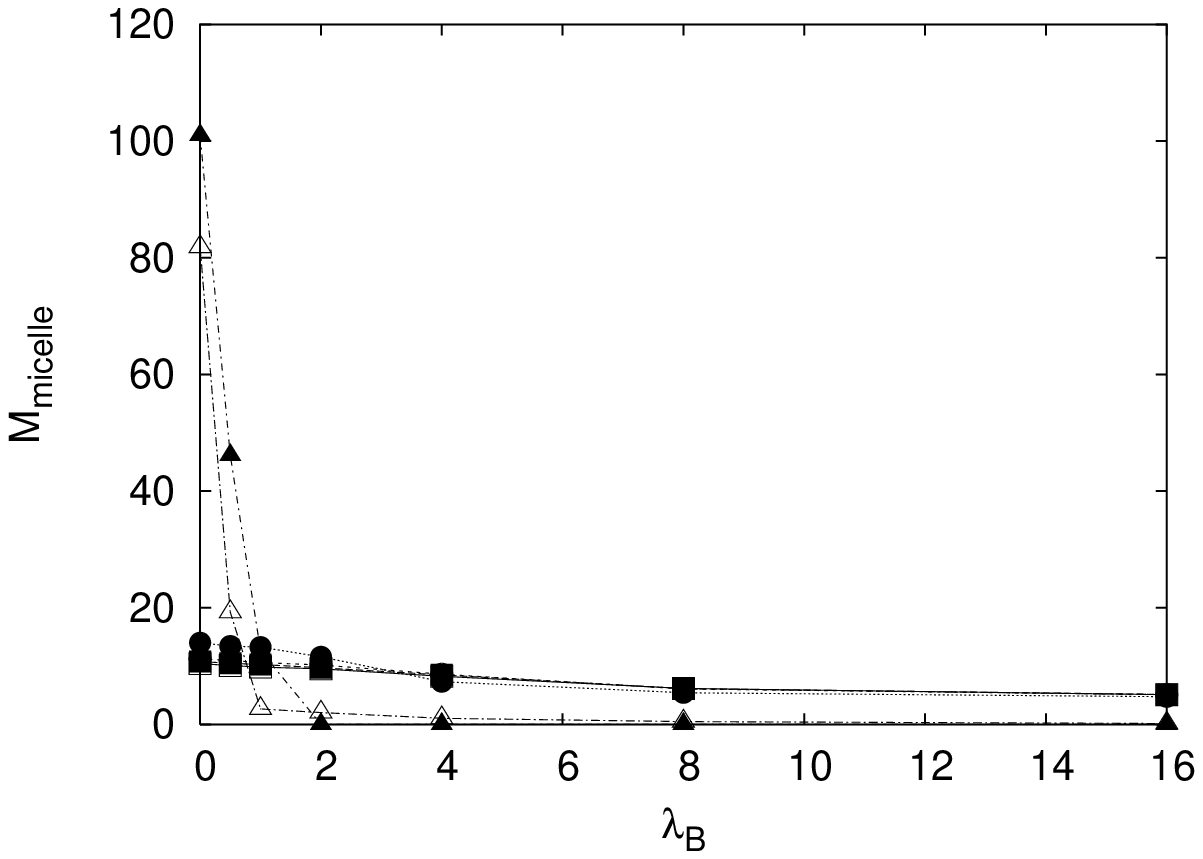}\\
\caption{\label{shapeUM1} \label{shapeUM2} 
(a) The total mass (number of beads) of the complex and (b) of the
largest detached micelle as function of the Coulomb strength
$\lambda_B$ for different values of the hydrophobicity parameter
$\varepsilon_{LJ}=0 (\square), 1 (\blacksquare), 2 ({\displaystyle
\circ}), 4 ({\displaystyle \bullet}), 6 (\vartriangle), 8
(\blacktriangle)$ from above.}
\end{figure}
The sizes of the complex and of the
largest detached micelle are shown in Fig. \ref{shapeUM1}
 where we plot the masses of these aggregates as
function of the Coulomb parameter $\lambda_B$ for different values of
the hydrophobicity $\varepsilon_{LJ}$.  It is clearly seen from these
plots that a large detached micelle appears only for small Coulombic
strength $\lambda_B<1$ and high hydrophobic attraction
$\varepsilon_{LJ}>5$. For these high $\varepsilon_{LJ}>4$ the largest
detached micelle shrinks rapidly with increasing $\lambda_B$ while at the same
time the mass of the complex grows such that the surfactant is nearly
quantitatively condensed in the complex for $\lambda_B>1$. For smaller
hydrophobicity $\varepsilon_{LJ}<4$ on the other hand no larger
detached micelles appear at all while the mass of the complex grows
only moderately for increasing Coulombic strength $\lambda_B$. Thus,
while for high hydrophobicity the size of the largest detached micelle is
inversely related to the size of the complex, the absence of large
micelles for smaller hydrophobicity may be explained with their
instability due to the Coulombic repulsion between the charged
surfactants in combination with entropy.  As shown in the appendix the
total binding energy of the surfactant tails in the micelle grows with
proportionally to its mass $M_{\rm mic}$ with a surface correction
proportional to $M^{2/3}_{\rm mic}$. The electrostatic energy of the
surfactant heads on the other hand is proportional to $M^{5/3}_{\rm
mic}$ while the entropy term behaves like $M_{\rm mic}\ln M_{\rm
mic}$. The qualitative behavior of these three terms together with the
approximate prefactors derived in the appendix may explain that for the
finite number of available surfactants the formation of large 
detached micelles is unfavorable for $\epsilon_{LJ}<6$ and even for
$\epsilon_{LJ}=6,8$ the mass of the largest detached micelle 
is significantly smaller than the total mass $M_{mic}^{max}=160$ of 
available surfactant. 
\subsection{Shapes of the complex}
In the case of complex formation we characterize the shape of
the complex in terms of its matrix of inertia
\begin{equation}
\Theta_{\alpha\beta} = \frac{1}{M}\sum_{j\in C}
({\bf r}_j^2\delta_{\alpha\beta} - r_j^{(\alpha)} r_j^{(\beta)})
\equiv\overline{ {r}^2\delta_{\alpha\beta} - r^{(\alpha)} r^{(\beta)} }
\end{equation}
where the sum is over all $M$ beads $j$ that are part of the complex and
${\bf r}_j=(r_j^{(1)},r_j^{(2)},r_j^{(3)})$ is the position of bead $j$
with respect to the center of mass of the complex.
\begin{figure}
\includegraphics[width=80mm]{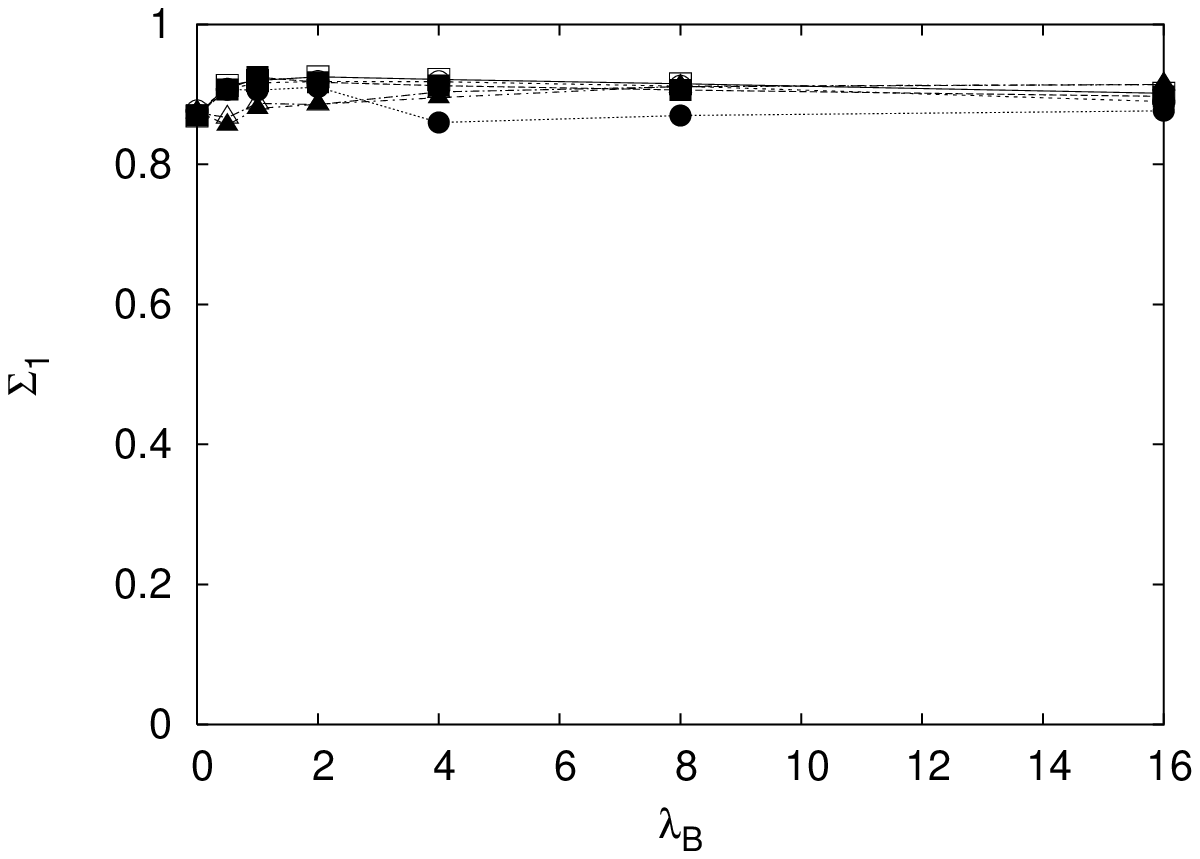}\\
\includegraphics[width=80mm]{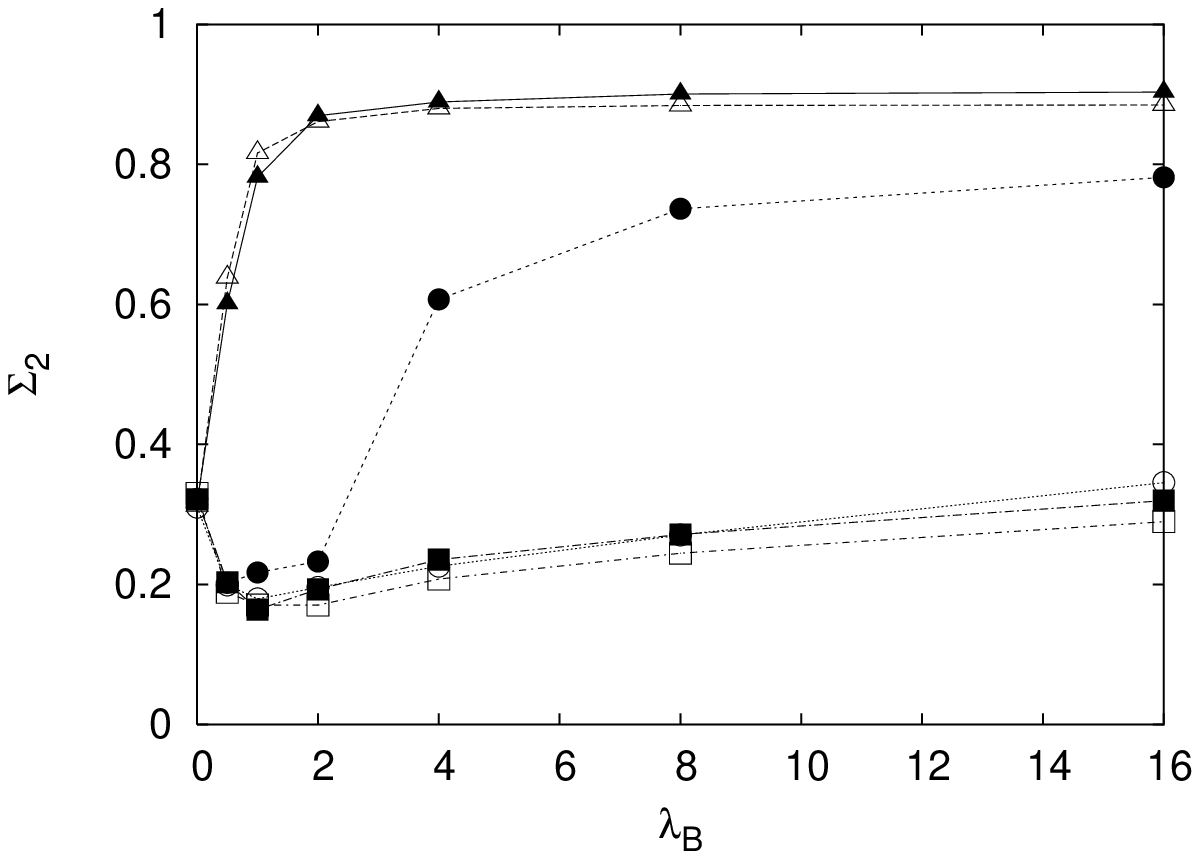}\\
\caption{\label{shapeUS1}\label{shapeUS2}  
(a) The ratio $\Sigma_1=\vartheta_y/\vartheta_x$ of the second largest and
largest principal moments of inertia and (b) the ratio
$\Sigma_2=\vartheta_z/\vartheta_x$ of the smallest and the largest 
principal moments of inertia as function of the Coulomb strength
$\lambda_B$ for different values of the hydrophobicity parameter
$\varepsilon_{LJ}=0 (\square), 1 (\blacksquare), 2 ({\displaystyle
\circ}), 4 ({\displaystyle \bullet}), 6 (\vartriangle), 8
(\blacktriangle)$ from below.}
\end{figure}
Diagonalizing $\Theta_{\alpha\beta}$ results in the three eigenvalues
$\vartheta_x\geq \vartheta_y\geq \vartheta_z$ with respective axes
$x,y,z$ of the principal moments of inertia with the center of mass 
as the origin. Note that in these coordinates one has e.g.
\begin{equation}
\vartheta_z=\overline{ x^2 + y^2 }
\end{equation}
such that the radius $R_G$ of gyration of the complex is given by
$R_G^2=(\vartheta_x+\vartheta_y+\vartheta_z)/2=\frac{1}{2}{\rm Tr} \Theta$.
The
relative values of the principal moments determine the global shape of
the complex. The averaged ratios 
$\Sigma_1=\langle\vartheta_y/\vartheta_x\rangle$ and 
$\Sigma_2=\langle\vartheta_z/\vartheta_x\rangle$ are plotted as functions of 
the Coulomb strength $\lambda_B$ for different hydrophobicities
$\varepsilon_{LJ}$ in Fig. \ref{shapeUS1}.  
These serve as indicators for the transition of the
complex from a cylindrical shape to an almost spherical one. The
cylindrical shape is characterized by two large approximately equal
inertia, i.e.  $\Sigma_1\approx 1$ and a third smaller moment such
that $\Sigma_2\approx 0.2...0.3$. For an elongated cylindrical shape
the $z$-axis corresponding to the smallest moment $\vartheta_z$ is the
symmetry axis. The spherical shapes in turn are characterized by three
almost equal moments of inertia with $\Sigma_2\approx \Sigma_1\approx
1$. We observe the following - see Fig. \ref{shapeUS1}: 
At $\lambda_B=0,\varepsilon_{LJ}=0$ only the PE chain is part of the 'complex'.
The corresponding values we find for the neutral chain are 
$\Sigma_1=\langle\vartheta_y/\vartheta_x\rangle=0.87$
and  $\Sigma_2=\langle\vartheta_z/\vartheta_x\rangle=0.34$. These may be 
compared with $\langle\vartheta_y\rangle/\langle\vartheta_x\rangle=0.87$
and  $\langle\vartheta_z\rangle/\langle\vartheta_x\rangle=0.27$ calculated
from data found for self avoiding walks \cite{Zifferer-1998}.
Note that the relation $\langle\vartheta_z/\vartheta_x\rangle>
\langle\vartheta_z\rangle/\langle\vartheta_x\rangle$ is in accordance
with the Schwartz inequality for averages.
For weak hydrophobicity $\varepsilon_{LJ}<3$ the shape of the complex remains cylindrical even for high Coulomb strength, while for strong hydrophobicity
$\varepsilon_{LJ}>5$ the transition from cylindrical to spherical shape
occurs already for small Coulomb strengths $\lambda_B \approx 1$.
For the intermediate hydrophobicity $\lambda_B =4$ though, we observe
a competition between the formation of the cylindrical bottle brush
and the spherical micellar complex. This fact is expressed by the 
non-monotonic behavior of the shape parameter $\Sigma_2$ as 
a function of $\lambda_B$. 
While for small $\lambda_B$ the cylindrical shape wins and  $\Sigma_2$
decreases with respect to the value of the neutral situation, 
a transition occurs between $\lambda_B=2$  and   $\lambda_B=4$ where
the shape parameter indicates a transition towards the spherical micellar
complex, with $\Sigma_2$ increasing to $0.6$ and further towards $0.8$ 
for $\lambda_B>8$.
\begin{figure}
\includegraphics[width=80mm]{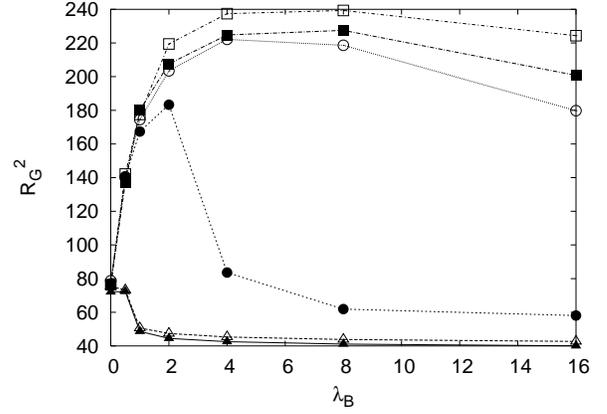}\\
\caption{\label{shapeUS0} The radius of gyration of the complex
as function of the Coulomb strength $\lambda_B$ for different values
of the hydrophobicity parameter 
$\varepsilon_{LJ}=0 (\square), 1 (\blacksquare), 2 ({\displaystyle \circ}),
4 ({\displaystyle \bullet}), 6 (\vartriangle), 8 (\blacktriangle)$ from above.}
\end{figure}
To verify the transition we analyze 
the internal structure of these intermediate states in more detail 
below. A similar division into strong, weak and intermediate hydrophobicity
we find for the behavior of the square radius of gyration $R_G^2$
which is plotted as function of $\lambda_B$ in Fig. \ref{shapeUS0}.
For small hydrophobicity  $\varepsilon_{LJ}<3$ the behavior of 
$R_G^2$ parallels that of the square end-to-end distance $R^2$
as described above for the case $\varepsilon_{LJ}=0$.
For high hydrophobicity $\varepsilon_{LJ}>5$ the radius of gyration of the complex shrinks already for small  $\lambda_B \approx 1$ to values
corresponding to compact micellar complexes. Again, the intermediate case
 $\varepsilon_{LJ} =4$ behaves special: For small Coulomb strengths 
$\lambda_B<3$ the radius grows due to the stretching of the 
polyelectrolyte chain
and only for higher $\lambda_B>6$ it again shrinks below the polymer coil
value but still remains well separated from the high hydrophobicity case.
\subsection{Internal structure of the complex}
While the overall shape of the complex is characterized by the
inertia matrix, its internal structure can be explored by either the
internal monomer-monomer correlations or by the density distributions
of the different complexes in a given reference frame. We have used
both approaches to identify the key features of the self-assembled
structures. To verify the idea of the molecular bottle brush like
structures in the case of low hydrophobicity $\varepsilon_{LJ}$ and high
Coulomb strength $\lambda_B$, we have measured in particular the
correlation between the polyelectrolyte monomers and the surfactant
tail ends.  
\begin{figure}
\includegraphics[width=80mm]{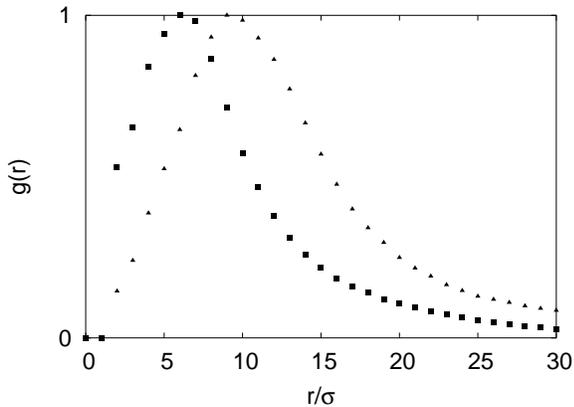}\\
\caption{\label{fig5}\label{g_of_r}
The pair correlation function $g(r)$ for pairs of a polyelectrolyte
monomer and a surfactant tail monomer at Bjerrum length
$\lambda_B=8\sigma$. The functions are scaled to a maximum value of
$1$.  The maximal number of tail ends is found at $r=6\sigma$ from the
polyelectrolyte backbone in the case of $n=5(\blacksquare)$ monomer 
surfactants. For $n=10 (\blacktriangle)$ this most probable distance 
is $r=9\sigma$.}
\end{figure}
As shown in Fig . \ref{g_of_r} using also data of
ref. \cite{Ferber-2003} we indeed find that the
surfactant tail end is found in this situation with highest
probability at a distance from the polyelectrolyte chain that
increases with the length of the tail in accordance with the bottle
brush picture.  To monitor, on the other hand, the internal
reorganization along with the transition from the cylindrical
bottle-brush to the spherical micellar aggregate, we chose as a
reference frame a coordinate system defined by the directions of the
principle moments of inertia of the complex with the origin at its
center of mass.  We define the $x,y,z$-directions such that the
corresponding moments are ordered by $\vartheta_x\geq\vartheta_y\geq
\vartheta_z $. As mentioned above,
the $z$-axis then is the symmetry axis for the
averaged cylindrical structures. 

\begin{figure}
\includegraphics[width=80mm]{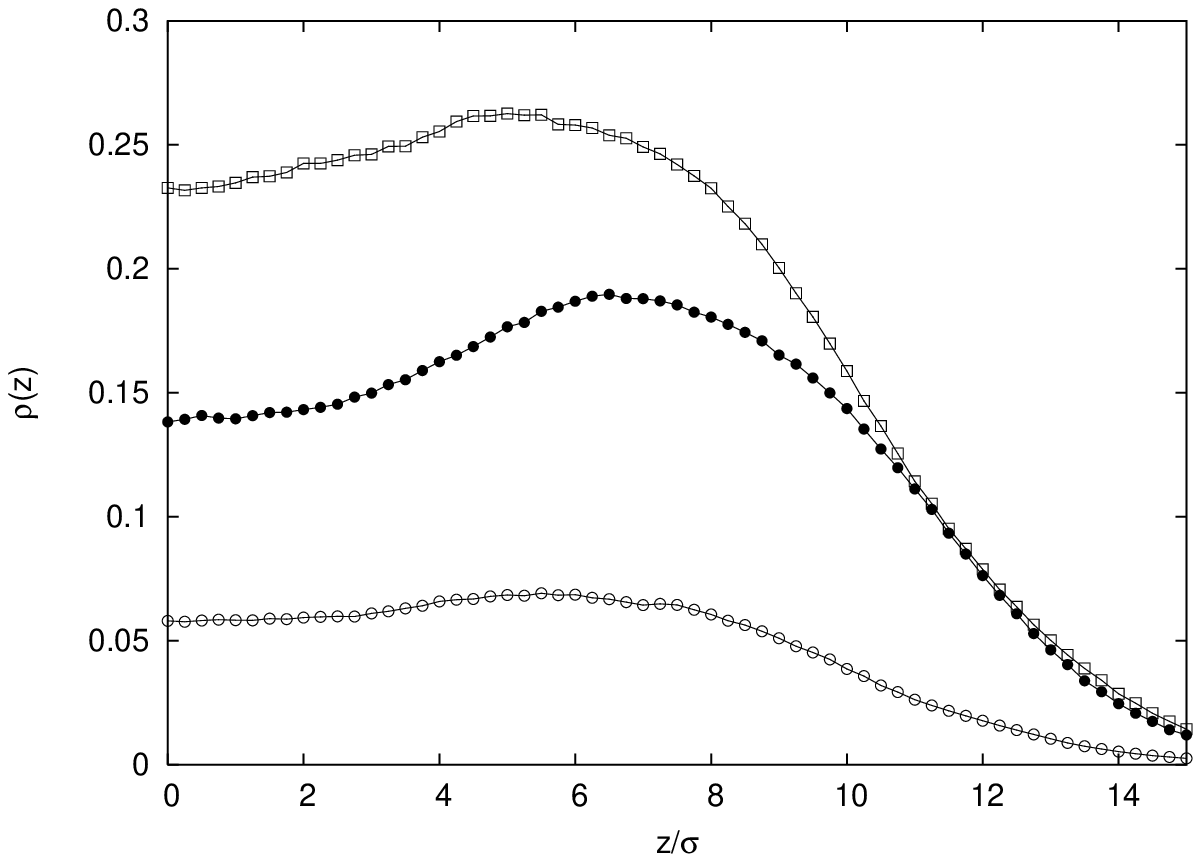}\\
\includegraphics[width=80mm]{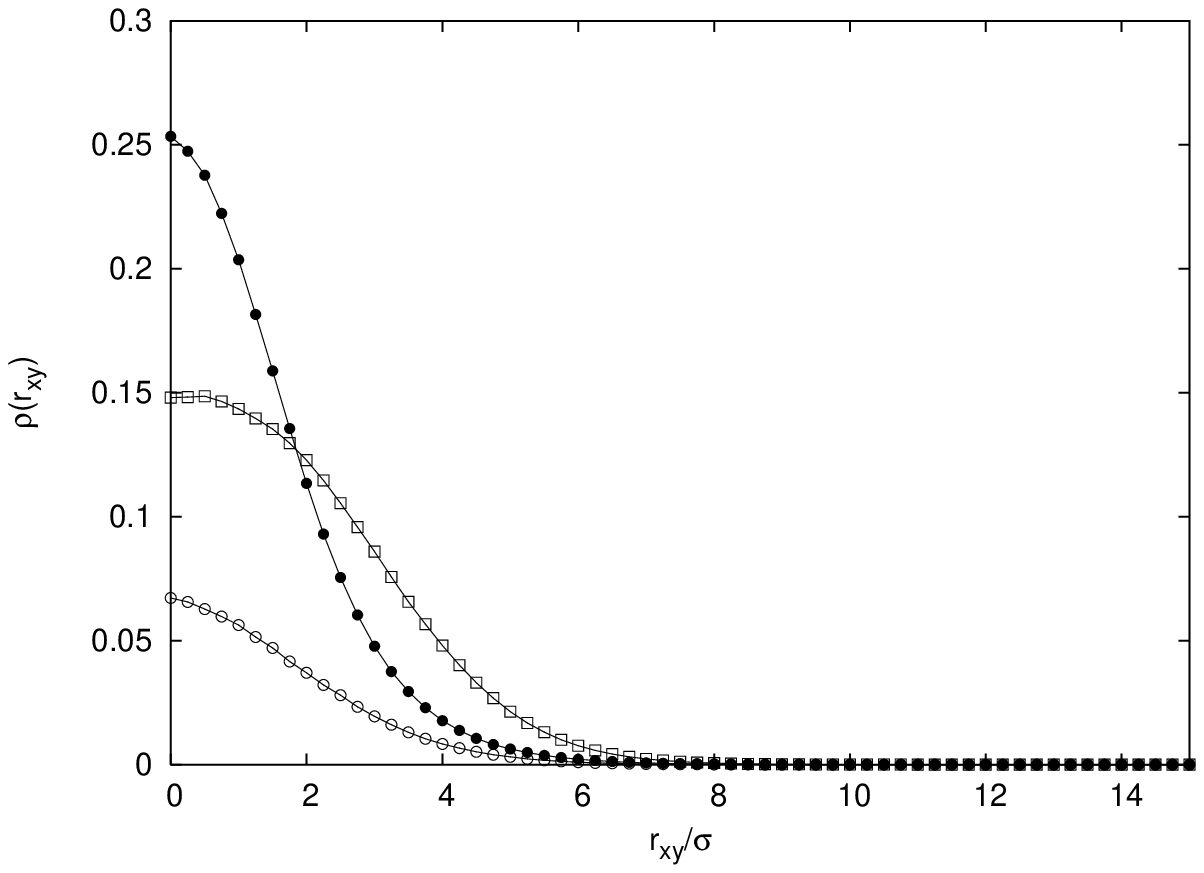}\\
\caption{\label{rho_of_r_z-04}\label{rho_of_r_xy-04}
 Density distributions of monomers 
in directions (a) along the $z$-axis and (b) 
perpendicular to the $z$-axis for $\lambda_B=0$, $\varepsilon_{LJ}=4$
for polyelectrolyte beads (${\displaystyle \bullet}$),
 surfactant heads (${\displaystyle \circ}$) and tail monomers ($\square$).}
\end{figure}
\begin{figure}
\includegraphics[width=80mm]{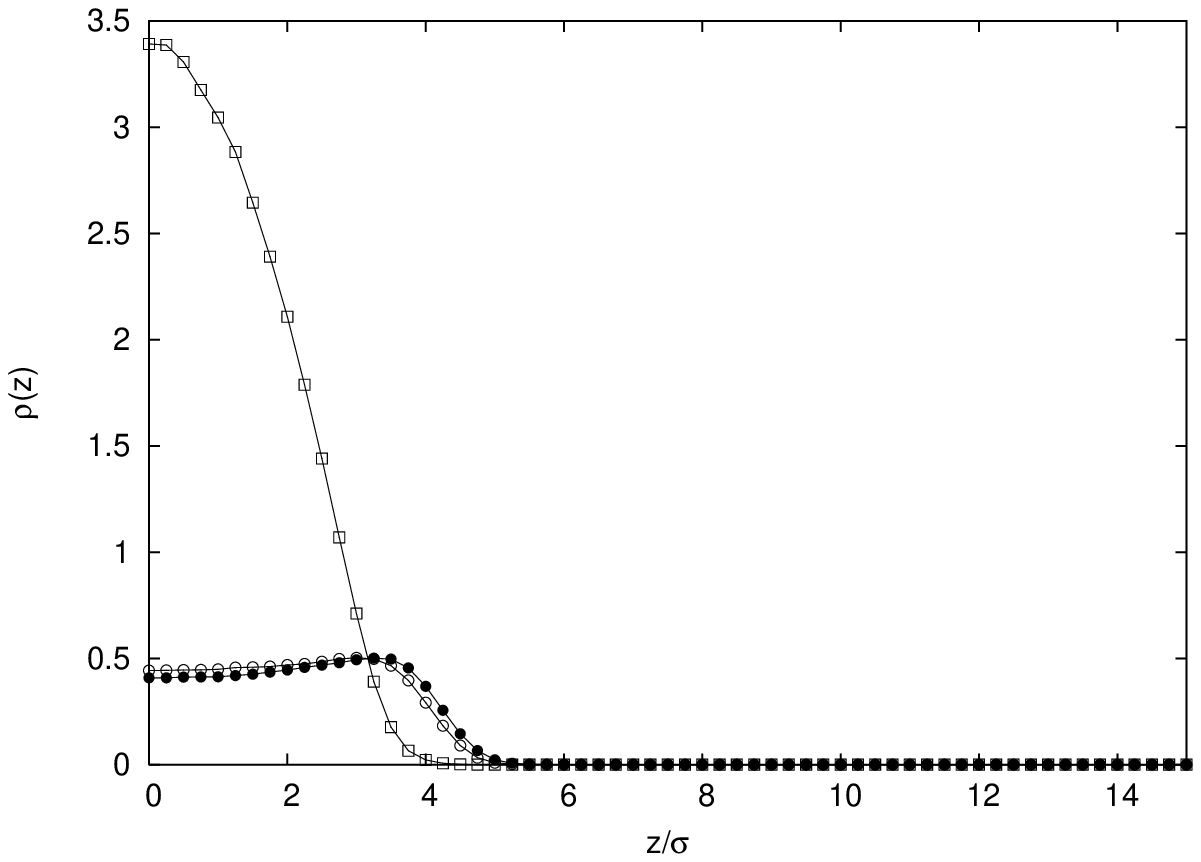}\\
\includegraphics[width=80mm]{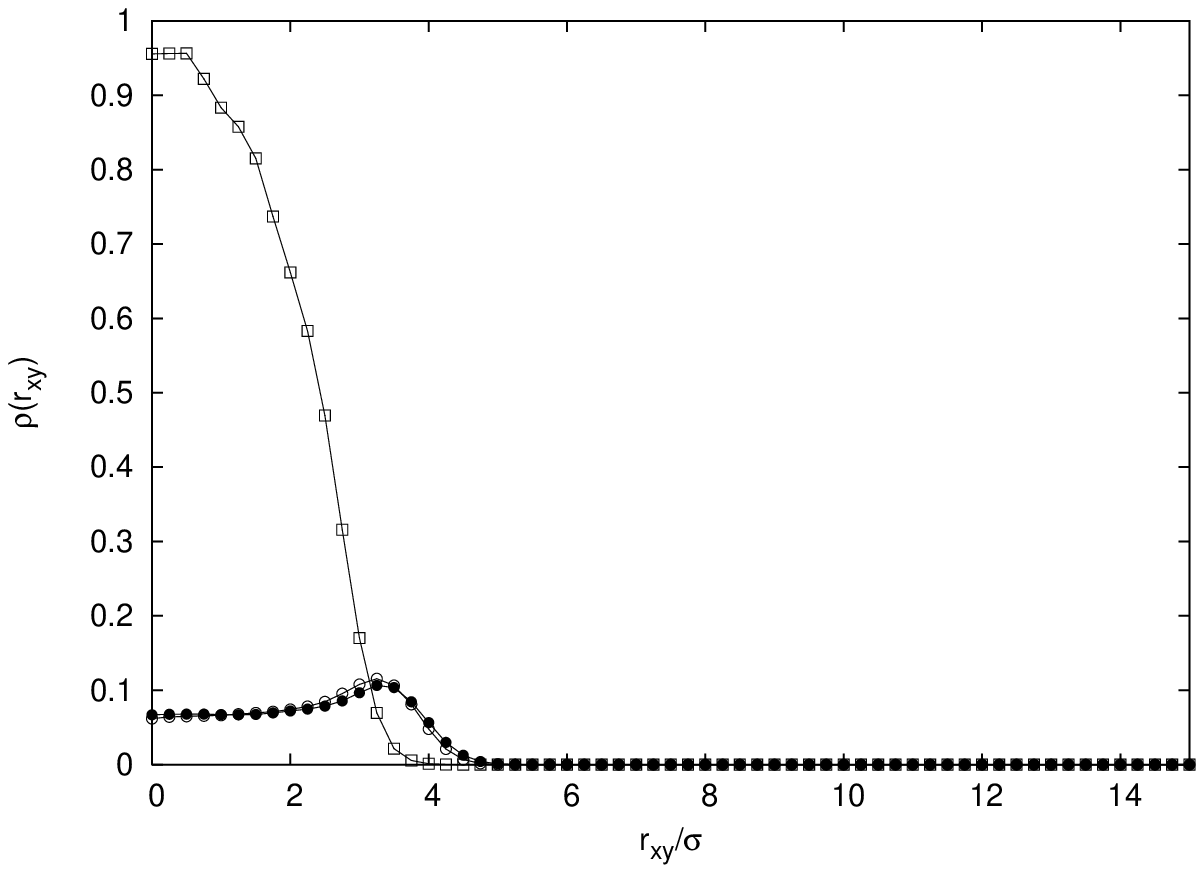}\\
\includegraphics[width=80mm]{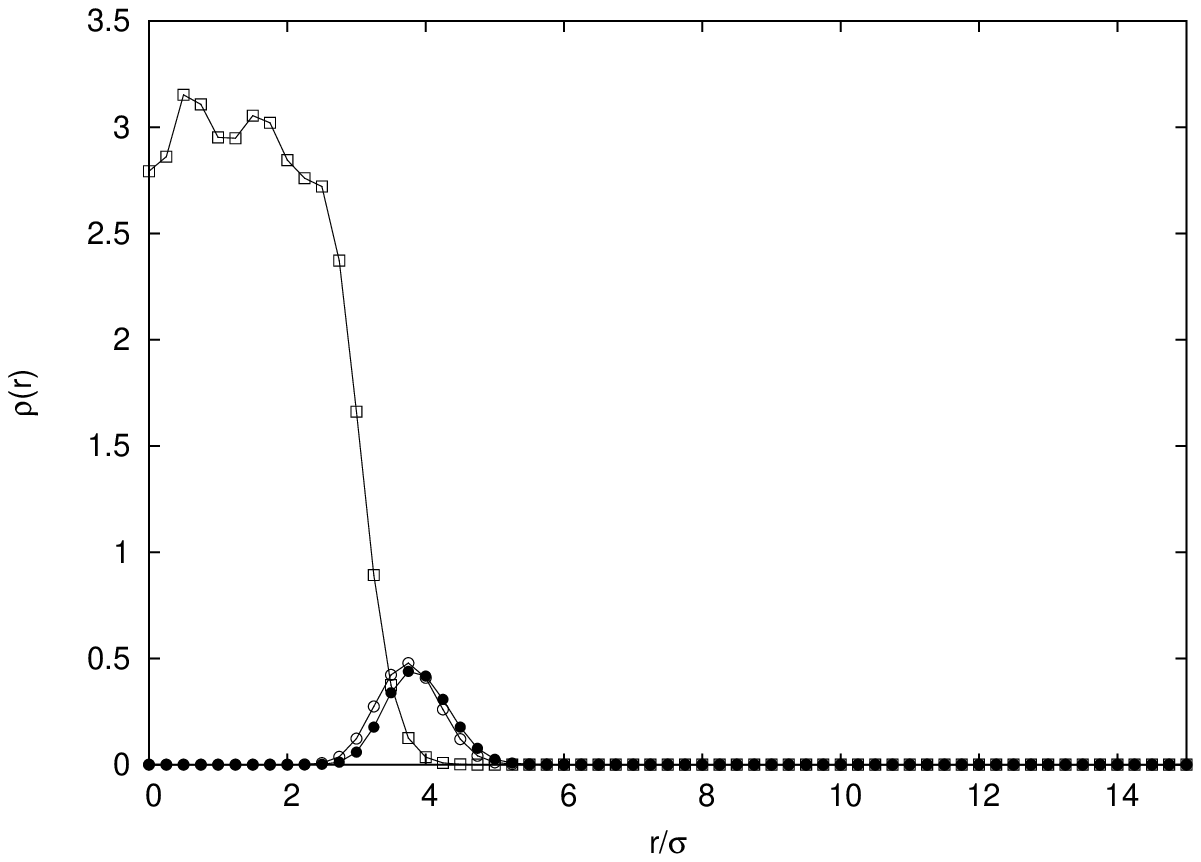}\\
\caption{\label{rho_of_r_z-06u04}
\label{rho_of_r_xy-06u04}
\label{rho_of_r-06u04}
 Density distributions of monomers 
in directions (a) along the $z$-axis, 
(b) perpendicular to the $z$-axis,
and (c) as function of the distance from the center of mass
for $\lambda_B=16$, $\varepsilon_{LJ}=8$
for polyelectrolyte beads (${\displaystyle \bullet}$),
 surfactant heads (${\displaystyle \circ}$) and tail monomers ($\square$).}
\end{figure}
\begin{figure}
\includegraphics[width=80mm]{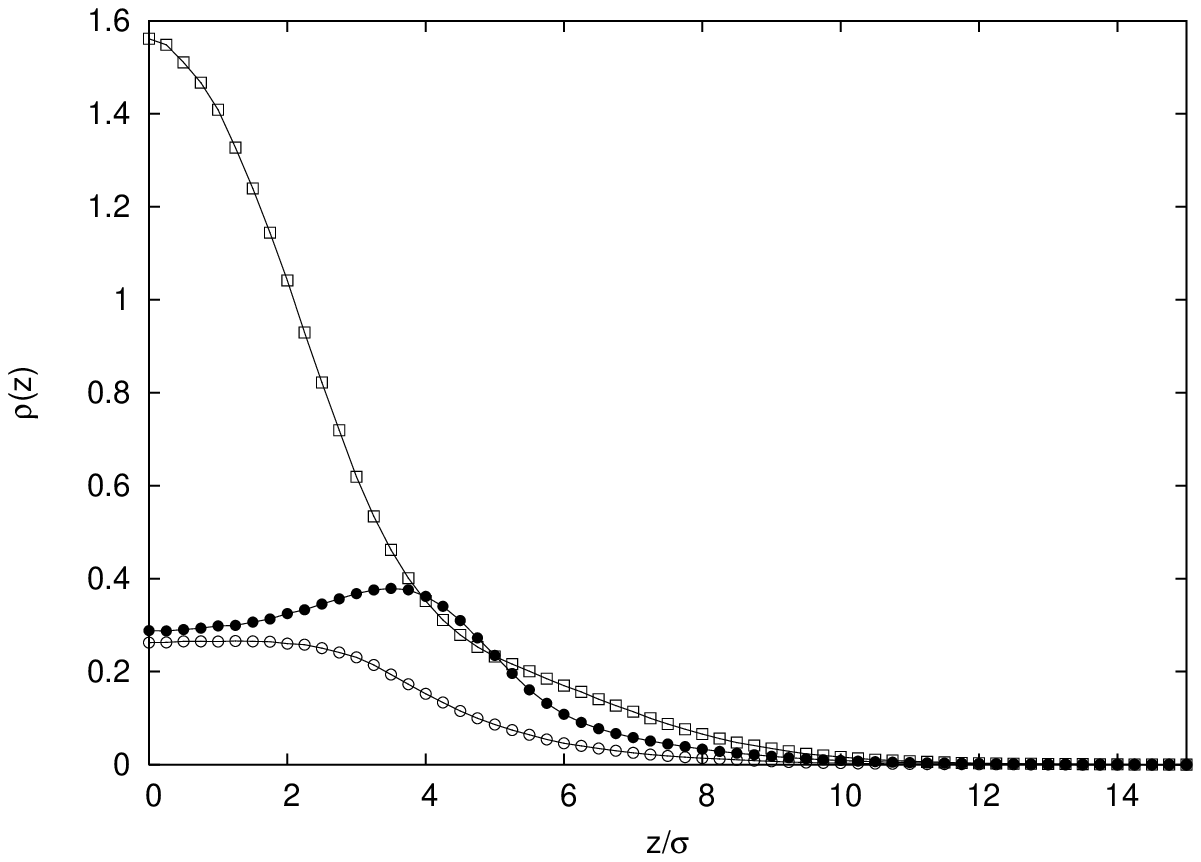}\\
\includegraphics[width=80mm]{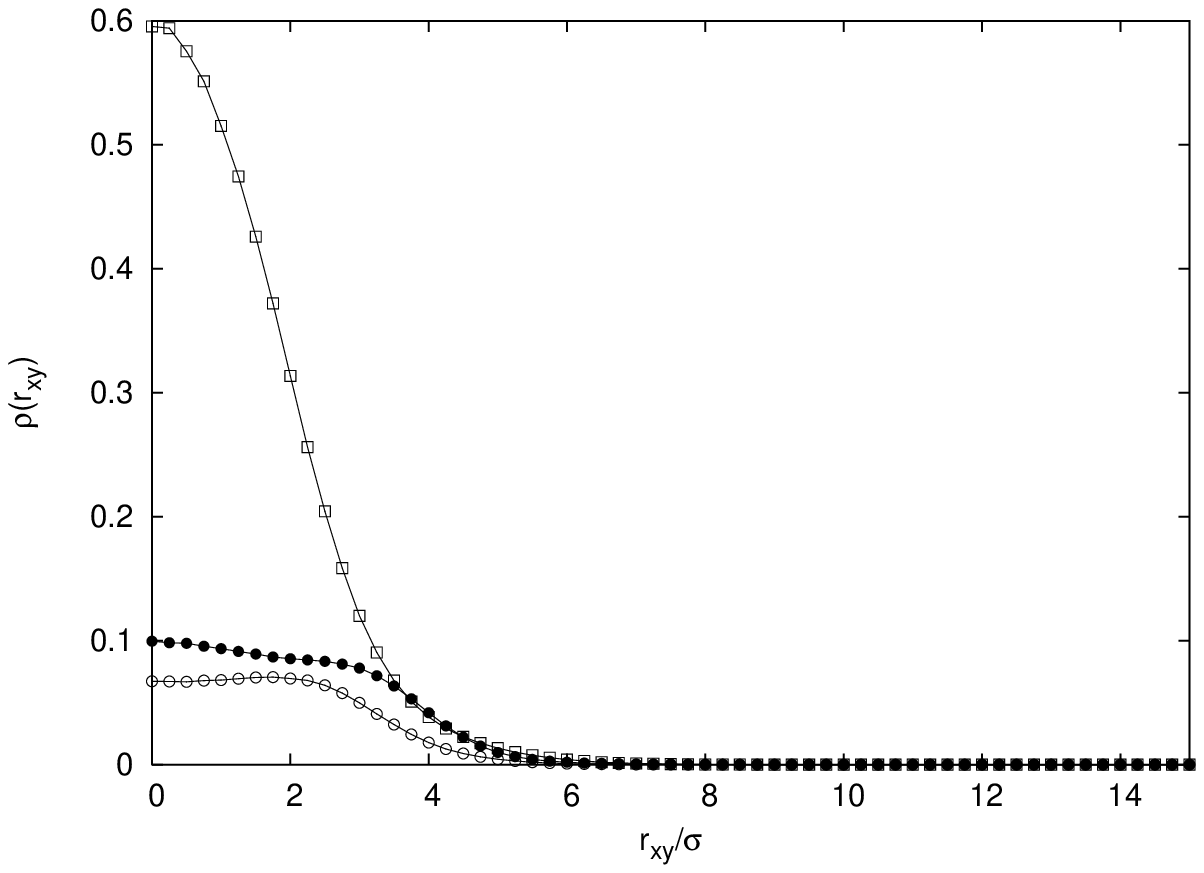}\\
\includegraphics[width=80mm]{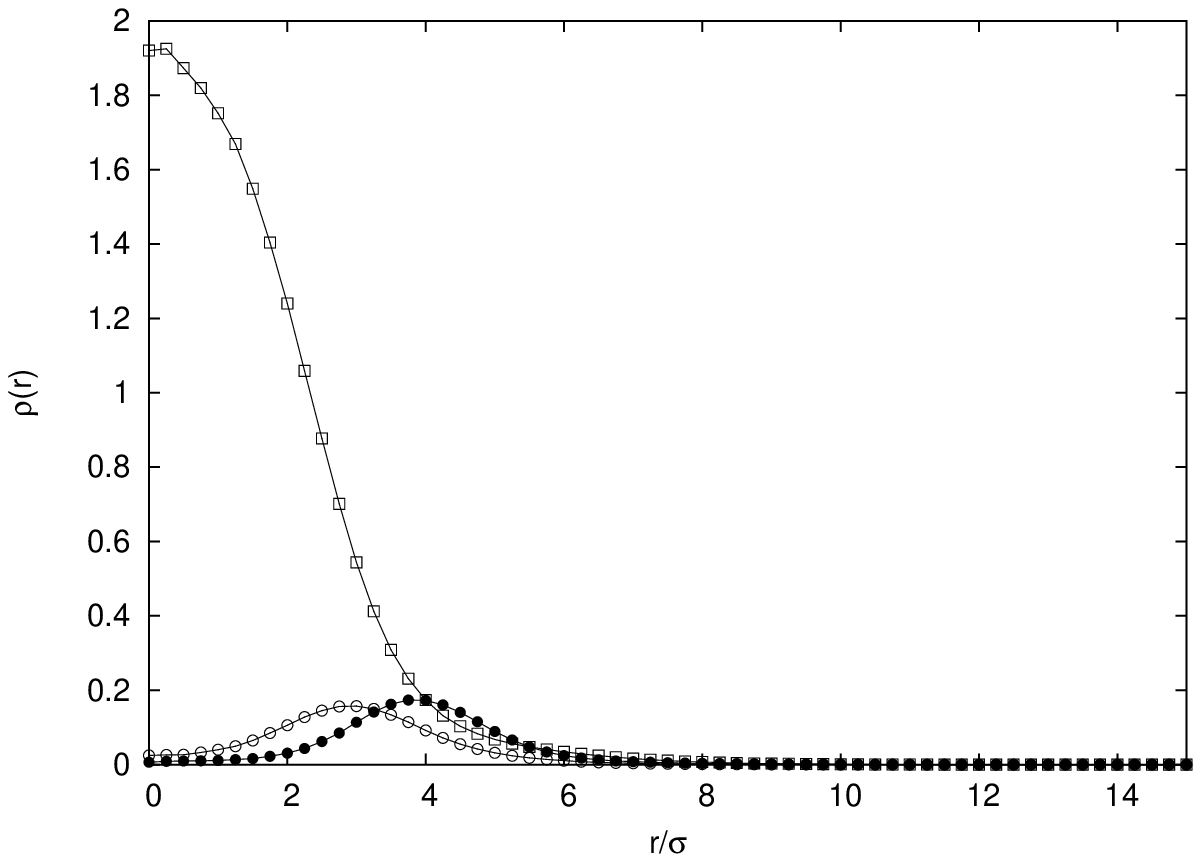}\\
\caption{\label{rho_of_r_z-04u03}
\label{rho_of_r_xy-04u03}
\label{rho_of_r-04u03}
 Density distributions of monomers 
in directions (a) along the $z$-axis, 
(b) perpendicular to the $z$-axis,
and (c) as function of the distance from the center of mass
for $\lambda_B=4$, $\varepsilon_{LJ}=4$
for polyelectrolyte beads (${\displaystyle \bullet}$),
 surfactant heads (${\displaystyle \circ}$) and tail monomers ($\square$).}
\end{figure}

In Figs. 
\ref{rho_of_r_z-04}-\ref{rho_of_r-04u03} we show
exemplarily for typical conformations of the complex the density
distributions measured in this coordinate system.  For vanishing
hydrophobicity $\varepsilon_{LJ}=0$ and $\lambda_B=4$ the
corresponding distributions are shown in
Figs. \ref{rho_of_r_xy-04}.  The densities as function
of the distance $r_{xy}$ from the $z$-axis as plotted in
Fig. \ref{rho_of_r_xy-04}(a) show clearly that the polyelectrolyte monomers
and the surfactant heads are localized near the symmetry ($z$)-axis 
while the density of the surfactant-tail monomers surpasses that of the former
two species at larger distances from the central axis.
The distributions measured along the $z$-axis on the other hand show the 
remarkable feature of maxima near the ends of the cylinder.
Surprising as it may seem, this effect is even exhibited by the equilibrium 
coil of a neutral polymer chain and in particular in our simulation
 for vanishing interactions.

The spherical micellar complexes, on the other hand, which we observe
for sufficiently large parameter values of $\lambda_B$ and
$\varepsilon_{LJ}$ are characterized by density distributions along
and perpendicular to the $z$-axis that are nearly
identical - see Fig. \ref{rho_of_r_xy-06u04}.
Only the different
differential volumes used for measurement lead to deviations.
A more natural variable in the spherical case is
of course the distance $r$ from the center of mass. The density
$\rho(r)$ as displayed in Fig. \ref{rho_of_r-06u04} parallels the
corresponding results of Wallin and Linse obtained by a self
consistent spherical lattice field approach \cite{Wallin-1998}: The
core of the micelle is formed by almost densely packed surfactant
tails surrounded by surfactant heads that stick out of the surface and
another layer that contains the polyelectrolyte chain.

A transitional structure found for the values  $\lambda_B=4$ and
$\varepsilon_{LJ}=4$ of the interactions near
to the boundary between the two states  is investigated in Fig.  
\ref{rho_of_r-04u03}. Although the structure
is much more loosely bound it rather parallels the micellar aggregate
displaying however a less pronounced layering in the spherical density
distribution, see Fig. \ref{rho_of_r-04u03}. The comparison of 
Figs.  \ref{rho_of_r_xy-04u03} (a) and (b) however shows
that the distributions display deviations from spherical symmetry.
Nonetheless, the present data indicates that the transition between the
two states of the complex is rather sharp.
\section{Conclusions}\label{V}
In extensive MC simulations we have investigated the behavior of a
single polyelectrolyte chain that interacts with oppositely charged
surfactant molecules taking into account both long range Coulomb and
short range hydrophobic interactions.  Exploring the parameter space
in both of these interactions we have established a ``phase diagram''
for this system and several distinct regions (``phases'') with
characteristic structural behavior of the system as sketched in
Fig. \ref{phasesketch}: for weak interactions both constituents behave
nearly independently and the PE chain has a possibly stretched coil
conformation.  For high Coulomb strength but weak hydrophobicity we
observe the formation of a complex with a cylindrical bottle brush
structure in which the PE chain is stretched along the cylinder axis.
If both interactions are strong, an inverted spherical micellar
complex structure emerges with the SF tails constituting the core of a
spherical micelle and the SF heads and the PE chain confined to the
surface of this sphere.  For intermediate values of the hydrophobicity
there is a competition between the two structurally different
complexation modes resulting in a non-trivial behavior of the shape
as function of the Coulomb strength.
For high hydrophobicity, but low Coulomb interaction 
the surfactant forms micelles to which the PE chain is only loosely
attached. 
As we have reported in Ref.\cite{Ferber-2003} complexation in this
system is in general weakened by the addition of salt. 
 Our results can in principle be
verified in experiments by systematically varying the surfactant tail
hydrophobicity and the surfactant/polyelectrolyte
charge and/or the dielectric permittivity of the solvent.

A number of extensions of our present approach are possible and would
be interesting to follow. Of particular interest would be a study of a
multi PE-chain system. For small numbers of chains a corresponding
simulation appears to be feasible using our present approach. One
might expect to observe in such a case a competition between the
cooperative assembly of a multichain complex and single chain
complexation.  At high densities in systems with many chains one
expects the formation of layered structures with many different
possible phases\cite{Thuene2}.  
However, high density and a large number of chains
are out of reach of the present approach, due to the requirement of
calculating the long range interaction between any pair of charged
beads and the slowing down of the simulation by the caging effect at
high densities. To simulate such systems either one has to release the
hard sphere non-penetrability \cite{Groot-2003} or resort to lattice
methods and simulate only with a screened shorter range Coulomb
interaction.

Other situations that can lead to significant changes in the behavior
might be found for a choice of different valency of the SF heads and
the PE chains to study over-charging effects of the polyelectrolyte
for high valency salt ions \cite{Messina-2002b}. Also, in general,
introducing neutral beads to the PE chain and varying the
distribution of charges along the chain may lead to significant
effects. In addition, the chain stiffness has been shown to be an important
parameter. For a single micelle modeled as a charged sphere, many
recent studies have revealed different topologies of the defect
depending on the chain stiffness
\cite{Wallin-1996,Gurovich-1999,Mateescu-1999,Netz-1999,Park-1999,Kunze-2000,Nguyen-2000,Welch-2000,Schiessel-2001,Jonsson-2001,Jonsson-2001a,Akinchina-2002,Chodanowski-2001,Chodanowski-2001a,Brynda-2002,Keren-2002}.

\section{Acknowledgments}\label{VI}

We thank A. F. Th\"unemann, O. Farago, H. Diamant and R. Messina for helpful
discussions.  This work was supported by the Deutsche
Forschungsgemeinschaft through grant No. LO 418/7.

\section{Appendix}
A crude argument for the stability of micelles built from
our model surfactants can be derived from the following considerations:
We assume that the core of a spheric micelle is given by $m_t$ tail
monomers at a packing fraction $\eta$ such that the core radius
$r_C$ is given by 
\begin{equation}
  r_C(m_t) = \frac{\sigma}{2}\left(\frac{m_t}{\eta}\right)^{1/3} .
\end{equation} 
The number of beads in the surface layer is then given by
\begin{equation}
  m_\sigma(m_t) = [r_C^3(m_t) - (r_C(m_t)- \frac{\sigma}{2})^3]\eta .
\end{equation}
Assuming that each bead inside the core has 12 neighbors at an optimal
distance (i.e. at the minimum of the potential $V_{LJ}$), while
the surface beads have only 6 neighbors we find the total 
Lennard-Jones energy of the micelle as 
\begin{equation} \label{E-LJ}
 \frac{1}{k_BT}E_{LJ}(m_t)=-(m_t-m_\sigma(m_t)/2)\frac{3}{2}\varepsilon_{LJ} .
\end{equation}
The total Coulomb energy of the SF heads distributed on the micellar
surface may be derived from the numerical solution of the Thomson problem
for distributing $k$ charges on a spherical surface which is given
by the fit formula \cite{Morris-1996}
\begin{equation}\label{E-Coulomb}
 \frac{1}{k_BT}E_{C}(k,m_t)=\frac{\lambda_B}{r_C(m_t)}\frac{k^2}{2}
(1-a k^{-1/2} + b k^{-3/2}),
\end{equation}
with $a=1.10461$ and $b=0.137$.
Finally, we may estimate the relevant entropy term by assuming that for every
molecule it is given by restricting it to the volume of the micelle.
This volume entropy term is thus
\begin{equation}\label{F-entropy}
\frac{1}{k_B}S(k,m_t)=k \ln(r(m_t)^3/\Omega)
\end{equation}
where $\Omega$ is the total system volume.  For our system of 32
surfactant molecules with one charged bead and four tail beads we have
$m_t=4k$ and a minimum of the free energy $F(k,m_t)$ with
\begin{equation}\label{F-free}
\frac{1}{k_BT} F(k,m_t)=E_C(k,m_t)+E_{LJ}(m_t)-S(k,m_t)
\end{equation}
at $n\neq 0$ exists
for a given Coulomb interaction only for sufficiently large
hydrophobicity.
Note, however that our estimate is too crude to give more than an
order of magnitude result showing that $\varepsilon_{LJ}$ should be at
least of the order of $\lambda_B$ and of the order of $\varepsilon_{LJ}\sim 1$
for  $\lambda_B=0$ to find micellar aggregation for our case of
surfactant length $n=5$, i.e. $m_t=4k$ and the given overall number of
surfactant molecules $k\leq 32$.
\bibliographystyle{apsrev}
\bibliography{pe04}

\begin{thebibliography}{73}
\expandafter\ifx\csname natexlab\endcsname\relax\def\natexlab#1{#1}\fi
\expandafter\ifx\csname bibnamefont\endcsname\relax
  \def\bibnamefont#1{#1}\fi
\expandafter\ifx\csname bibfnamefont\endcsname\relax
  \def\bibfnamefont#1{#1}\fi
\expandafter\ifx\csname citenamefont\endcsname\relax
  \def\citenamefont#1{#1}\fi
\expandafter\ifx\csname url\endcsname\relax
  \def\url#1{\texttt{#1}}\fi
\expandafter\ifx\csname urlprefix\endcsname\relax\def\urlprefix{URL }\fi
\providecommand{\bibinfo}[2]{#2}
\providecommand{\eprint}[2][]{\url{#2}}

\bibitem[{\citenamefont{Kwak}(1998)}]{Kwak-1998}
\bibinfo{author}{\bibfnamefont{J.~C.~T.} \bibnamefont{Kwak}},
  \emph{\bibinfo{title}{Polymer-Surfactant Systems}}
  (\bibinfo{publisher}{Marcel Dekker}, \bibinfo{address}{New York},
  \bibinfo{year}{1998}).

\bibitem[{\citenamefont{Antonietti et~al.}(1997)\citenamefont{Antonietti,
  Burger, and Th\"unemann}}]{Antonietti-1997}
\bibinfo{author}{\bibfnamefont{M.}~\bibnamefont{Antonietti}},
  \bibinfo{author}{\bibfnamefont{C.}~\bibnamefont{Burger}}, \bibnamefont{and}
  \bibinfo{author}{\bibfnamefont{A.}~\bibnamefont{Th\"unemann}},
  \bibinfo{journal}{Trends in Polymer Science} \textbf{\bibinfo{volume}{5}},
  \bibinfo{pages}{262} (\bibinfo{year}{1997}).

\bibitem[{\citenamefont{Hansson and Lindman}(1996)}]{Hansson-1996}
\bibinfo{author}{\bibfnamefont{P.}~\bibnamefont{Hansson}} \bibnamefont{and}
  \bibinfo{author}{\bibfnamefont{B.}~\bibnamefont{Lindman}},
  \bibinfo{journal}{Current Opinion in Colloid and Interface Science 1996}
  \textbf{\bibinfo{volume}{1}}, \bibinfo{pages}{604} (\bibinfo{year}{1996}).

\bibitem[{\citenamefont{Sokolov et~al.}(1998)\citenamefont{Sokolov, Yeh,
  Khokhlov, Grinberg, and Chu}}]{Sokolov-1998}
\bibinfo{author}{\bibfnamefont{E.}~\bibnamefont{Sokolov}},
  \bibinfo{author}{\bibfnamefont{F.}~\bibnamefont{Yeh}},
  \bibinfo{author}{\bibfnamefont{A.}~\bibnamefont{Khokhlov}},
  \bibinfo{author}{\bibfnamefont{V.~Y.} \bibnamefont{Grinberg}},
  \bibnamefont{and} \bibinfo{author}{\bibfnamefont{B.}~\bibnamefont{Chu}},
  \bibinfo{journal}{J. Phys. Chem. B} \textbf{\bibinfo{volume}{102}},
  \bibinfo{pages}{7091} (\bibinfo{year}{1998}).

\bibitem[{\citenamefont{Kosmella et~al.}(1998)\citenamefont{Kosmella, Kotz,
  Shirahama, and Liu}}]{Kosmella_1998}
\bibinfo{author}{\bibfnamefont{S.}~\bibnamefont{Kosmella}},
  \bibinfo{author}{\bibfnamefont{J.}~\bibnamefont{Kotz}},
  \bibinfo{author}{\bibfnamefont{K.}~\bibnamefont{Shirahama}},
  \bibnamefont{and} \bibinfo{author}{\bibfnamefont{J.}~\bibnamefont{Liu}},
  \bibinfo{journal}{J. Phys. Chem. B} \textbf{\bibinfo{volume}{102}},
  \bibinfo{pages}{6459} (\bibinfo{year}{1998}).

\bibitem[{\citenamefont{Claesson et~al.}(1998)\citenamefont{Claesson, Fielden,
  Dedinaite, Brown, and Fundin}}]{Claesson_1998}
\bibinfo{author}{\bibfnamefont{P.~M.} \bibnamefont{Claesson}},
  \bibinfo{author}{\bibfnamefont{M.~L.} \bibnamefont{Fielden}},
  \bibinfo{author}{\bibfnamefont{A.}~\bibnamefont{Dedinaite}},
  \bibinfo{author}{\bibfnamefont{W.}~\bibnamefont{Brown}}, \bibnamefont{and}
  \bibinfo{author}{\bibfnamefont{J.}~\bibnamefont{Fundin}},
  \bibinfo{journal}{J. Phys. Chem. B} \textbf{\bibinfo{volume}{102}},
  \bibinfo{pages}{1270} (\bibinfo{year}{1998}).

\bibitem[{\citenamefont{Tsianou and Alexandridis}(1999)}]{Tsianou-1999}
\bibinfo{author}{\bibfnamefont{M.}~\bibnamefont{Tsianou}} \bibnamefont{and}
  \bibinfo{author}{\bibfnamefont{P.}~\bibnamefont{Alexandridis}},
  \bibinfo{journal}{Langmuir} \textbf{\bibinfo{volume}{15}},
  \bibinfo{pages}{8105} (\bibinfo{year}{1999}).

\bibitem[{\citenamefont{Th\"unemann et~al.}(2000)\citenamefont{Th\"unemann,
  Kubowicz, and Pietsch}}]{Thunemann-2000a}
\bibinfo{author}{\bibfnamefont{A.~F.} \bibnamefont{Th\"unemann}},
  \bibinfo{author}{\bibfnamefont{S.}~\bibnamefont{Kubowicz}}, \bibnamefont{and}
  \bibinfo{author}{\bibfnamefont{U.}~\bibnamefont{Pietsch}},
  \bibinfo{journal}{Langmuir} \textbf{\bibinfo{volume}{16}},
  \bibinfo{pages}{8562} (\bibinfo{year}{2000}).

\bibitem[{\citenamefont{Dias et~al.}(2000)\citenamefont{Dias, Mel'nikov,
  Lindman, and Miguel}}]{Dias-2000}
\bibinfo{author}{\bibfnamefont{R.}~\bibnamefont{Dias}},
  \bibinfo{author}{\bibfnamefont{S.}~\bibnamefont{Mel'nikov}},
  \bibinfo{author}{\bibfnamefont{B.}~\bibnamefont{Lindman}}, \bibnamefont{and}
  \bibinfo{author}{\bibfnamefont{M.~G.} \bibnamefont{Miguel}},
  \bibinfo{journal}{Langmuir} \textbf{\bibinfo{volume}{16}},
  \bibinfo{pages}{9577} (\bibinfo{year}{2000}).

\bibitem[{\citenamefont{Babak et~al.}(2000)\citenamefont{Babak, Merkovich,
  Desbrieres, and Rinaudo}}]{Babak-2000}
\bibinfo{author}{\bibfnamefont{V.~G.} \bibnamefont{Babak}},
  \bibinfo{author}{\bibfnamefont{E.~A.} \bibnamefont{Merkovich}},
  \bibinfo{author}{\bibfnamefont{J.}~\bibnamefont{Desbrieres}},
  \bibnamefont{and} \bibinfo{author}{\bibfnamefont{M.}~\bibnamefont{Rinaudo}},
  \bibinfo{journal}{Polymer Bulletin} \textbf{\bibinfo{volume}{45}},
  \bibinfo{pages}{77} (\bibinfo{year}{2000}).

\bibitem[{\citenamefont{Liao and Higgins}(2001)}]{Liao-2001}
\bibinfo{author}{\bibfnamefont{X.~M.} \bibnamefont{Liao}} \bibnamefont{and}
  \bibinfo{author}{\bibfnamefont{D.~A.} \bibnamefont{Higgins}},
  \bibinfo{journal}{Langmuir} \textbf{\bibinfo{volume}{17}},
  \bibinfo{pages}{6051} (\bibinfo{year}{2001}).

\bibitem[{\citenamefont{Guan et~al.}(2001)\citenamefont{Guan, Cao, Peng, Xu,
  and Chen}}]{Guan-2001}
\bibinfo{author}{\bibfnamefont{Y.}~\bibnamefont{Guan}},
  \bibinfo{author}{\bibfnamefont{Y.~P.} \bibnamefont{Cao}},
  \bibinfo{author}{\bibfnamefont{Y.~X.} \bibnamefont{Peng}},
  \bibinfo{author}{\bibfnamefont{J.}~\bibnamefont{Xu}}, \bibnamefont{and}
  \bibinfo{author}{\bibfnamefont{A.~S.~C.} \bibnamefont{Chen}},
  \bibinfo{journal}{Chemical Communications} \textbf{\bibinfo{volume}{17}},
  \bibinfo{pages}{1694} (\bibinfo{year}{2001}).

\bibitem[{\citenamefont{Taylor and Thomas}(2002)}]{Taylor-2002}
\bibinfo{author}{\bibfnamefont{D.~J.~F.} \bibnamefont{Taylor}}
  \bibnamefont{and} \bibinfo{author}{\bibfnamefont{R.~K.}
  \bibnamefont{Thomas}}, \bibinfo{journal}{Langmuir}
  \textbf{\bibinfo{volume}{18}}, \bibinfo{pages}{4748} (\bibinfo{year}{2002}).

\bibitem[{\citenamefont{Hansson et~al.}(2002)\citenamefont{Hansson, Schneider,
  and Lindman}}]{Hansson-2002}
\bibinfo{author}{\bibfnamefont{P.}~\bibnamefont{Hansson}},
  \bibinfo{author}{\bibfnamefont{S.}~\bibnamefont{Schneider}},
  \bibnamefont{and} \bibinfo{author}{\bibfnamefont{B.}~\bibnamefont{Lindman}},
  \bibinfo{journal}{J. Chem. Phys. B} \textbf{\bibinfo{volume}{106}},
  \bibinfo{pages}{9777} (\bibinfo{year}{2002}).

\bibitem[{\citenamefont{Guillot et~al.}(2003)\citenamefont{Guillot, Loughlin,
  Jain, Delsanti, and Langevin}}]{Guillot-2003}
\bibinfo{author}{\bibfnamefont{S.}~\bibnamefont{Guillot}},
  \bibinfo{author}{\bibfnamefont{D.~M.} \bibnamefont{Loughlin}},
  \bibinfo{author}{\bibfnamefont{N.}~\bibnamefont{Jain}},
  \bibinfo{author}{\bibfnamefont{M.}~\bibnamefont{Delsanti}}, \bibnamefont{and}
  \bibinfo{author}{\bibfnamefont{D.}~\bibnamefont{Langevin}},
  \bibinfo{journal}{J. Phys.: Condensed Matter} \textbf{\bibinfo{volume}{15}},
  \bibinfo{pages}{S219} (\bibinfo{year}{2003}).

\bibitem[{\citenamefont{von Ferber and L\"owen}(2003)}]{Ferber-2003}
\bibinfo{author}{\bibfnamefont{C.}~\bibnamefont{von Ferber}} \bibnamefont{and}
  \bibinfo{author}{\bibfnamefont{H.}~\bibnamefont{L\"owen}},
  \bibinfo{journal}{J. Chem. Phys.} \textbf{\bibinfo{volume}{118}},
  \bibinfo{pages}{10774} (\bibinfo{year}{2003}).

\bibitem[{\citenamefont{Wallin and Linse}(1998)}]{Wallin-1998}
\bibinfo{author}{\bibfnamefont{T.}~\bibnamefont{Wallin}} \bibnamefont{and}
  \bibinfo{author}{\bibfnamefont{P.}~\bibnamefont{Linse}},
  \bibinfo{journal}{Langmuir} \textbf{\bibinfo{volume}{14}},
  \bibinfo{pages}{2940} (\bibinfo{year}{1998}).

\bibitem[{\citenamefont{Shirahama et~al.}(1981)\citenamefont{Shirahama, Yuasa,
  and Sugimoto}}]{Shira1}
\bibinfo{author}{\bibfnamefont{K.}~\bibnamefont{Shirahama}},
  \bibinfo{author}{\bibfnamefont{H.}~\bibnamefont{Yuasa}}, \bibnamefont{and}
  \bibinfo{author}{\bibfnamefont{S.}~\bibnamefont{Sugimoto}},
  \bibinfo{journal}{Bull. Chem. Soc. Jpn.} \textbf{\bibinfo{volume}{54}},
  \bibinfo{pages}{375} (\bibinfo{year}{1981}).

\bibitem[{\citenamefont{Shirahama and Tashiro}(1984)}]{Shira2}
\bibinfo{author}{\bibfnamefont{K.}~\bibnamefont{Shirahama}} \bibnamefont{and}
  \bibinfo{author}{\bibfnamefont{M.}~\bibnamefont{Tashiro}},
  \bibinfo{journal}{Bull. Chem. Soc. Jpn.} \textbf{\bibinfo{volume}{57}},
  \bibinfo{pages}{377} (\bibinfo{year}{1984}).

\bibitem[{\citenamefont{Sear}(1998)}]{Sear}
\bibinfo{author}{\bibfnamefont{R.}~\bibnamefont{Sear}}, \bibinfo{journal}{J.
  Phys.: Condensed Matter} \textbf{\bibinfo{volume}{10}}, \bibinfo{pages}{1677}
  (\bibinfo{year}{1998}).

\bibitem[{\citenamefont{Kuhn et~al.}(1998)\citenamefont{Kuhn, Levin, and
  Barbosa}}]{Kuhn-1998}
\bibinfo{author}{\bibfnamefont{P.~S.} \bibnamefont{Kuhn}},
  \bibinfo{author}{\bibfnamefont{Y.}~\bibnamefont{Levin}}, \bibnamefont{and}
  \bibinfo{author}{\bibfnamefont{M.~C.} \bibnamefont{Barbosa}},
  \bibinfo{journal}{Chemical Physics Letters} \textbf{\bibinfo{volume}{298}},
  \bibinfo{pages}{51} (\bibinfo{year}{1998}).

\bibitem[{\citenamefont{Kuhn et~al.}(2000)\citenamefont{Kuhn, Barbosa, and
  Levin}}]{Kuhn-2000}
\bibinfo{author}{\bibfnamefont{P.~S.} \bibnamefont{Kuhn}},
  \bibinfo{author}{\bibfnamefont{M.~C.} \bibnamefont{Barbosa}},
  \bibnamefont{and} \bibinfo{author}{\bibfnamefont{Y.}~\bibnamefont{Levin}},
  \bibinfo{journal}{Physica A} \textbf{\bibinfo{volume}{283}},
  \bibinfo{pages}{113} (\bibinfo{year}{2000}).

\bibitem[{\citenamefont{Silva et~al.}(2001)\citenamefont{Silva, Kuhn, and
  Lucena}}]{Silva-2001}
\bibinfo{author}{\bibfnamefont{M.~B.~A.} \bibnamefont{Silva}},
  \bibinfo{author}{\bibfnamefont{P.~S.} \bibnamefont{Kuhn}}, \bibnamefont{and}
  \bibinfo{author}{\bibfnamefont{L.~S.} \bibnamefont{Lucena}},
  \bibinfo{journal}{Physica A} \textbf{\bibinfo{volume}{296}},
  \bibinfo{pages}{31} (\bibinfo{year}{2001}).

\bibitem[{\citenamefont{Diamant and
  Andelman}(2000{\natexlab{a}})}]{Diamant-2000a}
\bibinfo{author}{\bibfnamefont{H.}~\bibnamefont{Diamant}} \bibnamefont{and}
  \bibinfo{author}{\bibfnamefont{D.}~\bibnamefont{Andelman}},
  \bibinfo{journal}{Phys. Rev. E} \textbf{\bibinfo{volume}{61}},
  \bibinfo{pages}{6740} (\bibinfo{year}{2000}{\natexlab{a}}).

\bibitem[{\citenamefont{Diamant and
  Andelman}(2000{\natexlab{b}})}]{Diamant-2000b}
\bibinfo{author}{\bibfnamefont{H.}~\bibnamefont{Diamant}} \bibnamefont{and}
  \bibinfo{author}{\bibfnamefont{D.}~\bibnamefont{Andelman}},
  \bibinfo{journal}{Macromolecules} \textbf{\bibinfo{volume}{33}},
  \bibinfo{pages}{8050} (\bibinfo{year}{2000}{\natexlab{b}}).

\bibitem[{\citenamefont{Wallin and Linse}(1996)}]{Wallin-1996}
\bibinfo{author}{\bibfnamefont{T.}~\bibnamefont{Wallin}} \bibnamefont{and}
  \bibinfo{author}{\bibfnamefont{P.}~\bibnamefont{Linse}},
  \bibinfo{journal}{Langmuir} \textbf{\bibinfo{volume}{12}},
  \bibinfo{pages}{305} (\bibinfo{year}{1996}).

\bibitem[{\citenamefont{Stevens and Kremer}(1995)}]{Kremer1}
\bibinfo{author}{\bibfnamefont{M.}~\bibnamefont{Stevens}} \bibnamefont{and}
  \bibinfo{author}{\bibfnamefont{K.}~\bibnamefont{Kremer}},
  \bibinfo{journal}{J. Chem. Phys.} \textbf{\bibinfo{volume}{103}},
  \bibinfo{pages}{1669} (\bibinfo{year}{1995}).

\bibitem[{\citenamefont{Winkler et~al.}(1998)\citenamefont{Winkler, Gold, and
  Reineker}}]{Winkler}
\bibinfo{author}{\bibfnamefont{R.~G.} \bibnamefont{Winkler}},
  \bibinfo{author}{\bibfnamefont{M.}~\bibnamefont{Gold}}, \bibnamefont{and}
  \bibinfo{author}{\bibfnamefont{P.}~\bibnamefont{Reineker}},
  \bibinfo{journal}{Phys. Rev. Letters} \textbf{\bibinfo{volume}{80}},
  \bibinfo{pages}{3731} (\bibinfo{year}{1998}).

\bibitem[{\citenamefont{Micka et~al.}(1999)\citenamefont{Micka, Holm, and
  Kremer}}]{Kremer2}
\bibinfo{author}{\bibfnamefont{U.}~\bibnamefont{Micka}},
  \bibinfo{author}{\bibfnamefont{C.}~\bibnamefont{Holm}}, \bibnamefont{and}
  \bibinfo{author}{\bibfnamefont{K.}~\bibnamefont{Kremer}},
  \bibinfo{journal}{Langmuir} \textbf{\bibinfo{volume}{15}},
  \bibinfo{pages}{4033} (\bibinfo{year}{1999}).

\bibitem[{\citenamefont{Brilliantov et~al.}(1998)\citenamefont{Brilliantov,
  Kuznetsov, and Klein}}]{Brilliant}
\bibinfo{author}{\bibfnamefont{N.~V.} \bibnamefont{Brilliantov}},
  \bibinfo{author}{\bibfnamefont{D.~V.} \bibnamefont{Kuznetsov}},
  \bibnamefont{and} \bibinfo{author}{\bibfnamefont{R.}~\bibnamefont{Klein}},
  \bibinfo{journal}{Phys. Rev. Letters} \textbf{\bibinfo{volume}{81}},
  \bibinfo{pages}{1433} (\bibinfo{year}{1998}).

\bibitem[{\citenamefont{Kuhn}(2002)}]{Kuhn-2002}
\bibinfo{author}{\bibfnamefont{P.~S.} \bibnamefont{Kuhn}},
  \bibinfo{journal}{Physica A} \textbf{\bibinfo{volume}{311}},
  \bibinfo{pages}{50} (\bibinfo{year}{2002}).

\bibitem[{\citenamefont{Gurovitch and Sens}(1999)}]{Gurovich-1999}
\bibinfo{author}{\bibfnamefont{E.}~\bibnamefont{Gurovitch}} \bibnamefont{and}
  \bibinfo{author}{\bibfnamefont{P.}~\bibnamefont{Sens}},
  \bibinfo{journal}{Phys. Rev. Letters} \textbf{\bibinfo{volume}{82}},
  \bibinfo{pages}{339} (\bibinfo{year}{1999}).

\bibitem[{\citenamefont{Mateescu et~al.}(1999)\citenamefont{Mateescu, Jeppesen,
  and Pincus}}]{Mateescu-1999}
\bibinfo{author}{\bibfnamefont{E.}~\bibnamefont{Mateescu}},
  \bibinfo{author}{\bibfnamefont{C.}~\bibnamefont{Jeppesen}}, \bibnamefont{and}
  \bibinfo{author}{\bibfnamefont{P.}~\bibnamefont{Pincus}},
  \bibinfo{journal}{Europhys. Letters} \textbf{\bibinfo{volume}{46}},
  \bibinfo{pages}{493} (\bibinfo{year}{1999}).

\bibitem[{\citenamefont{Netz and Joanny}(1999)}]{Netz-1999}
\bibinfo{author}{\bibfnamefont{R.~R.} \bibnamefont{Netz}} \bibnamefont{and}
  \bibinfo{author}{\bibfnamefont{J.-F.} \bibnamefont{Joanny}},
  \bibinfo{journal}{Macromolecules} \textbf{\bibinfo{volume}{32}},
  \bibinfo{pages}{9026} (\bibinfo{year}{1999}).

\bibitem[{\citenamefont{Park et~al.}(1999)\citenamefont{Park, Bruinsma, and
  Gelbart}}]{Park-1999}
\bibinfo{author}{\bibfnamefont{S.~Y.} \bibnamefont{Park}},
  \bibinfo{author}{\bibfnamefont{R.~F.} \bibnamefont{Bruinsma}},
  \bibnamefont{and} \bibinfo{author}{\bibfnamefont{W.~M.}
  \bibnamefont{Gelbart}}, \bibinfo{journal}{Europhys. Letters}
  \textbf{\bibinfo{volume}{46}}, \bibinfo{pages}{454} (\bibinfo{year}{1999}).

\bibitem[{\citenamefont{Kunze and Netz}(2000)}]{Kunze-2000}
\bibinfo{author}{\bibfnamefont{K.~K.} \bibnamefont{Kunze}} \bibnamefont{and}
  \bibinfo{author}{\bibfnamefont{R.~R.} \bibnamefont{Netz}},
  \bibinfo{journal}{Phys. Rev. Letters} \textbf{\bibinfo{volume}{85}},
  \bibinfo{pages}{4389} (\bibinfo{year}{2000}).

\bibitem[{\citenamefont{Nguyen and Shklovskii}(2000)}]{Nguyen-2000}
\bibinfo{author}{\bibfnamefont{T.~T.} \bibnamefont{Nguyen}} \bibnamefont{and}
  \bibinfo{author}{\bibfnamefont{B.~I.} \bibnamefont{Shklovskii}},
  \bibinfo{journal}{Physica A} \textbf{\bibinfo{volume}{293}},
  \bibinfo{pages}{324} (\bibinfo{year}{2000}).

\bibitem[{\citenamefont{Welch and Muthukumar}(2000)}]{Welch-2000}
\bibinfo{author}{\bibfnamefont{P.}~\bibnamefont{Welch}} \bibnamefont{and}
  \bibinfo{author}{\bibfnamefont{M.}~\bibnamefont{Muthukumar}},
  \bibinfo{journal}{Macromolecules} \textbf{\bibinfo{volume}{33}},
  \bibinfo{pages}{6159} (\bibinfo{year}{2000}).

\bibitem[{\citenamefont{Schiessel et~al.}(2001)\citenamefont{Schiessel,
  Bruinsma, and Gelbart}}]{Schiessel-2001}
\bibinfo{author}{\bibfnamefont{H.}~\bibnamefont{Schiessel}},
  \bibinfo{author}{\bibfnamefont{R.~F.} \bibnamefont{Bruinsma}},
  \bibnamefont{and} \bibinfo{author}{\bibfnamefont{W.~M.}
  \bibnamefont{Gelbart}}, \bibinfo{journal}{J. Chem. Phys.}
  \textbf{\bibinfo{volume}{115}}, \bibinfo{pages}{7245} (\bibinfo{year}{2001}).

\bibitem[{\citenamefont{Jonsson and Linse}(2001{\natexlab{a}})}]{Jonsson-2001}
\bibinfo{author}{\bibfnamefont{M.}~\bibnamefont{Jonsson}} \bibnamefont{and}
  \bibinfo{author}{\bibfnamefont{P.}~\bibnamefont{Linse}}, \bibinfo{journal}{J.
  Chem. Phys.} \textbf{\bibinfo{volume}{115}}, \bibinfo{pages}{3406}
  (\bibinfo{year}{2001}{\natexlab{a}}).

\bibitem[{\citenamefont{Jonsson and Linse}(2001{\natexlab{b}})}]{Jonsson-2001a}
\bibinfo{author}{\bibfnamefont{M.}~\bibnamefont{Jonsson}} \bibnamefont{and}
  \bibinfo{author}{\bibfnamefont{P.}~\bibnamefont{Linse}}, \bibinfo{journal}{J.
  Chem. Phys.} \textbf{\bibinfo{volume}{115}}, \bibinfo{pages}{10975}
  (\bibinfo{year}{2001}{\natexlab{b}}).

\bibitem[{\citenamefont{Akinchina and Linse}(2002)}]{Akinchina-2002}
\bibinfo{author}{\bibfnamefont{A.}~\bibnamefont{Akinchina}} \bibnamefont{and}
  \bibinfo{author}{\bibfnamefont{P.}~\bibnamefont{Linse}},
  \bibinfo{journal}{Macromolecules} \textbf{\bibinfo{volume}{35}},
  \bibinfo{pages}{5183} (\bibinfo{year}{2002}).

\bibitem[{\citenamefont{Chodanowski and
  Stoll}(2001{\natexlab{a}})}]{Chodanowski-2001}
\bibinfo{author}{\bibfnamefont{P.}~\bibnamefont{Chodanowski}} \bibnamefont{and}
  \bibinfo{author}{\bibfnamefont{S.}~\bibnamefont{Stoll}}, \bibinfo{journal}{J.
  Chem. Phys.} \textbf{\bibinfo{volume}{115}}, \bibinfo{pages}{4951}
  (\bibinfo{year}{2001}{\natexlab{a}}).

\bibitem[{\citenamefont{Chodanowski and
  Stoll}(2001{\natexlab{b}})}]{Chodanowski-2001a}
\bibinfo{author}{\bibfnamefont{P.}~\bibnamefont{Chodanowski}} \bibnamefont{and}
  \bibinfo{author}{\bibfnamefont{S.}~\bibnamefont{Stoll}},
  \bibinfo{journal}{Macromolecules} \textbf{\bibinfo{volume}{34}},
  \bibinfo{pages}{2320} (\bibinfo{year}{2001}{\natexlab{b}}).

\bibitem[{\citenamefont{Brynda et~al.}(2002)\citenamefont{Brynda, Chodanowski,
  and Stoll}}]{Brynda-2002}
\bibinfo{author}{\bibfnamefont{M.}~\bibnamefont{Brynda}},
  \bibinfo{author}{\bibfnamefont{P.}~\bibnamefont{Chodanowski}},
  \bibnamefont{and} \bibinfo{author}{\bibfnamefont{S.}~\bibnamefont{Stoll}},
  \bibinfo{journal}{Colloid Polym. Sci.} \textbf{\bibinfo{volume}{280}},
  \bibinfo{pages}{789} (\bibinfo{year}{2002}).

\bibitem[{\citenamefont{Keren et~al.}(2002)\citenamefont{Keren, Soen, Yoseph,
  Gilad, Braun, Silvan, , and Talmon}}]{Keren-2002}
\bibinfo{author}{\bibfnamefont{K.}~\bibnamefont{Keren}},
  \bibinfo{author}{\bibfnamefont{Y.}~\bibnamefont{Soen}},
  \bibinfo{author}{\bibfnamefont{G.~B.} \bibnamefont{Yoseph}},
  \bibinfo{author}{\bibfnamefont{R.}~\bibnamefont{Gilad}},
  \bibinfo{author}{\bibfnamefont{E.}~\bibnamefont{Braun}},
  \bibinfo{author}{\bibfnamefont{U.}~\bibnamefont{Silvan}}, , \bibnamefont{and}
  \bibinfo{author}{\bibfnamefont{Y.}~\bibnamefont{Talmon}},
  \bibinfo{journal}{Phys. Rev. Letters} \textbf{\bibinfo{volume}{89}},
  \bibinfo{pages}{088103} (\bibinfo{year}{2002}).

\bibitem[{\citenamefont{ten Brinke and Ikkala}(1997)}]{tenBrinke-1997}
\bibinfo{author}{\bibfnamefont{G.}~\bibnamefont{ten Brinke}} \bibnamefont{and}
  \bibinfo{author}{\bibfnamefont{O.}~\bibnamefont{Ikkala}},
  \bibinfo{journal}{Trends in Polymer Science} \textbf{\bibinfo{volume}{5}},
  \bibinfo{pages}{213} (\bibinfo{year}{1997}).

\bibitem[{\citenamefont{Ruokolainen et~al.}(1996)\citenamefont{Ruokolainen, ten
  Brinke, Ikkala, Torkkeli, and Serimaa}}]{Ruokolainen-1996}
\bibinfo{author}{\bibfnamefont{J.}~\bibnamefont{Ruokolainen}},
  \bibinfo{author}{\bibfnamefont{G.}~\bibnamefont{ten Brinke}},
  \bibinfo{author}{\bibfnamefont{O.}~\bibnamefont{Ikkala}},
  \bibinfo{author}{\bibfnamefont{M.}~\bibnamefont{Torkkeli}}, \bibnamefont{and}
  \bibinfo{author}{\bibfnamefont{R.}~\bibnamefont{Serimaa}},
  \bibinfo{journal}{Macromolecules} \textbf{\bibinfo{volume}{29}},
  \bibinfo{pages}{3409} (\bibinfo{year}{1996}).

\bibitem[{\citenamefont{Birshtein et~al.}(1987)\citenamefont{Birshtein,
  Borisov, Zhulina, Khokhlov, and Yurasova}}]{Birshtein-1987}
\bibinfo{author}{\bibfnamefont{T.~M.} \bibnamefont{Birshtein}},
  \bibinfo{author}{\bibfnamefont{O.~V.} \bibnamefont{Borisov}},
  \bibinfo{author}{\bibfnamefont{Y.~B.} \bibnamefont{Zhulina}},
  \bibinfo{author}{\bibfnamefont{A.~R.} \bibnamefont{Khokhlov}},
  \bibnamefont{and} \bibinfo{author}{\bibfnamefont{T.~A.}
  \bibnamefont{Yurasova}}, \bibinfo{journal}{Polym. Sci. USSR}
  \textbf{\bibinfo{volume}{29}}, \bibinfo{pages}{1293} (\bibinfo{year}{1987}).

\bibitem[{\citenamefont{Fredrickson}(1993)}]{Fredrickson-1993}
\bibinfo{author}{\bibfnamefont{G.~H.} \bibnamefont{Fredrickson}},
  \bibinfo{journal}{Macromolecules} \textbf{\bibinfo{volume}{26}},
  \bibinfo{pages}{2825} (\bibinfo{year}{1993}).

\bibitem[{\citenamefont{Rouault and Borisov}(1996)}]{Rouault-1996}
\bibinfo{author}{\bibfnamefont{Y.}~\bibnamefont{Rouault}} \bibnamefont{and}
  \bibinfo{author}{\bibfnamefont{O.~V.} \bibnamefont{Borisov}},
  \bibinfo{journal}{Macromolecules} \textbf{\bibinfo{volume}{29}},
  \bibinfo{pages}{2605} (\bibinfo{year}{1996}).

\bibitem[{\citenamefont{Khalatur et~al.}(2000)\citenamefont{Khalatur,
  Shirvanyanz, Starovoitova, and Khokhlov}}]{Khalatur-2000}
\bibinfo{author}{\bibfnamefont{P.~G.} \bibnamefont{Khalatur}},
  \bibinfo{author}{\bibfnamefont{D.~G.} \bibnamefont{Shirvanyanz}},
  \bibinfo{author}{\bibfnamefont{N.~Y.} \bibnamefont{Starovoitova}},
  \bibnamefont{and} \bibinfo{author}{\bibfnamefont{A.~R.}
  \bibnamefont{Khokhlov}}, \bibinfo{journal}{Macromolecular theory and
  simulations} \textbf{\bibinfo{volume}{9}}, \bibinfo{pages}{141}
  (\bibinfo{year}{2000}).

\bibitem[{\citenamefont{Hansen and L{\"o}wen}(2000)}]{Hansen-2000}
\bibinfo{author}{\bibfnamefont{J.-P.} \bibnamefont{Hansen}} \bibnamefont{and}
  \bibinfo{author}{\bibfnamefont{H.}~\bibnamefont{L{\"o}wen}},
  \bibinfo{journal}{Annual Reviews of Physical Chemistry}
  \textbf{\bibinfo{volume}{51}}, \bibinfo{pages}{209} (\bibinfo{year}{2000}).

\bibitem[{\citenamefont{Groot}(2000)}]{Groot-2000}
\bibinfo{author}{\bibfnamefont{R.~D.} \bibnamefont{Groot}},
  \bibinfo{journal}{Langmuir} \textbf{\bibinfo{volume}{16}},
  \bibinfo{pages}{7493} (\bibinfo{year}{2000}).

\bibitem[{\citenamefont{Rescic and Linse}(2000)}]{Rescic-2000}
\bibinfo{author}{\bibfnamefont{J.}~\bibnamefont{Rescic}} \bibnamefont{and}
  \bibinfo{author}{\bibfnamefont{P.}~\bibnamefont{Linse}}, \bibinfo{journal}{J.
  Phys. Chem. B} \textbf{\bibinfo{volume}{104}}, \bibinfo{pages}{7852}
  (\bibinfo{year}{2000}).

\bibitem[{\citenamefont{Wallin and Linse}(1997)}]{Wallin-1997}
\bibinfo{author}{\bibfnamefont{T.}~\bibnamefont{Wallin}} \bibnamefont{and}
  \bibinfo{author}{\bibfnamefont{P.}~\bibnamefont{Linse}}, \bibinfo{journal}{J.
  Phys. Chem. B} \textbf{\bibinfo{volume}{101}}, \bibinfo{pages}{5506}
  (\bibinfo{year}{1997}).

\bibitem[{\citenamefont{Winkler et~al.}(2002)\citenamefont{Winkler,
  Steinhauser, and Reineker}}]{Winkler-2002}
\bibinfo{author}{\bibfnamefont{R.~G.} \bibnamefont{Winkler}},
  \bibinfo{author}{\bibfnamefont{M.~O.} \bibnamefont{Steinhauser}},
  \bibnamefont{and} \bibinfo{author}{\bibfnamefont{P.}~\bibnamefont{Reineker}},
  \bibinfo{journal}{Phys. Rev. E} \textbf{\bibinfo{volume}{66}},
  \bibinfo{pages}{021802} (\bibinfo{year}{2002}).

\bibitem[{\citenamefont{Jusufi et~al.}(2002)\citenamefont{Jusufi, Likos, and
  L{\"o}wen}}]{Jusufi-2002}
\bibinfo{author}{\bibfnamefont{A.}~\bibnamefont{Jusufi}},
  \bibinfo{author}{\bibfnamefont{C.~N.} \bibnamefont{Likos}}, \bibnamefont{and}
  \bibinfo{author}{\bibfnamefont{H.}~\bibnamefont{L{\"o}wen}},
  \bibinfo{journal}{Phys. Rev. Letters} \textbf{\bibinfo{volume}{88}},
  \bibinfo{pages}{018301} (\bibinfo{year}{2002}).

\bibitem[{\citenamefont{Messina
  et~al.}(2002{\natexlab{a}})\citenamefont{Messina, Holm, and
  Kremer}}]{Messina-2002a}
\bibinfo{author}{\bibfnamefont{R.}~\bibnamefont{Messina}},
  \bibinfo{author}{\bibfnamefont{C.}~\bibnamefont{Holm}}, \bibnamefont{and}
  \bibinfo{author}{\bibfnamefont{K.}~\bibnamefont{Kremer}},
  \bibinfo{journal}{Phys. Rev. E} \textbf{\bibinfo{volume}{65}},
  \bibinfo{pages}{041805} (\bibinfo{year}{2002}{\natexlab{a}}).

\bibitem[{\citenamefont{Odijk}(1979)}]{Odijk-1979}
\bibinfo{author}{\bibfnamefont{T.}~\bibnamefont{Odijk}},
  \bibinfo{journal}{Macromolecules} \textbf{\bibinfo{volume}{12}},
  \bibinfo{pages}{688} (\bibinfo{year}{1979}).

\bibitem[{\citenamefont{Fixman and Skolnik}(1978)}]{Fixman-1978}
\bibinfo{author}{\bibfnamefont{M.}~\bibnamefont{Fixman}} \bibnamefont{and}
  \bibinfo{author}{\bibfnamefont{J.}~\bibnamefont{Skolnik}},
  \bibinfo{journal}{Macromolecules} \textbf{\bibinfo{volume}{11}},
  \bibinfo{pages}{863} (\bibinfo{year}{1978}).

\bibitem[{\citenamefont{deGennes et~al.}(1976)\citenamefont{deGennes, Pincus,
  and Velasco}}]{deGennes-1976}
\bibinfo{author}{\bibfnamefont{P.~G.} \bibnamefont{deGennes}},
  \bibinfo{author}{\bibfnamefont{P.}~\bibnamefont{Pincus}}, \bibnamefont{and}
  \bibinfo{author}{\bibfnamefont{R.}~\bibnamefont{Velasco}},
  \bibinfo{journal}{J. Phys. (Paris)} \textbf{\bibinfo{volume}{37}},
  \bibinfo{pages}{1461} (\bibinfo{year}{1976}).

\bibitem[{\citenamefont{Barrat and Joanny}(1993)}]{Barrat-1993}
\bibinfo{author}{\bibfnamefont{J.-L.} \bibnamefont{Barrat}} \bibnamefont{and}
  \bibinfo{author}{\bibfnamefont{J.-F.} \bibnamefont{Joanny}},
  \bibinfo{journal}{Europhys. Lett.} \textbf{\bibinfo{volume}{3}},
  \bibinfo{pages}{343} (\bibinfo{year}{1993}).

\bibitem[{\citenamefont{Connolly et~al.}(2003)\citenamefont{Connolly,
  Timoshenko, and Kuznetsov}}]{Connolly-2003}
\bibinfo{author}{\bibfnamefont{R.}~\bibnamefont{Connolly}},
  \bibinfo{author}{\bibfnamefont{E.~G.} \bibnamefont{Timoshenko}},
  \bibnamefont{and} \bibinfo{author}{\bibfnamefont{Y.~A.}
  \bibnamefont{Kuznetsov}}, \bibinfo{journal}{J. Chem. Phys.}
  \textbf{\bibinfo{volume}{119}}, \bibinfo{pages}{8736} (\bibinfo{year}{2003}).

\bibitem[{\citenamefont{Timoshenko and Kuznetsov}(2000)}]{Timoshenko-2000}
\bibinfo{author}{\bibfnamefont{E.~G.} \bibnamefont{Timoshenko}}
  \bibnamefont{and} \bibinfo{author}{\bibfnamefont{Y.~A.}
  \bibnamefont{Kuznetsov}}, \bibinfo{journal}{J. Chem. Phys.}
  \textbf{\bibinfo{volume}{112}}, \bibinfo{pages}{8163} (\bibinfo{year}{2000}).

\bibitem[{\citenamefont{Lekner}(1991)}]{Lekner-1991}
\bibinfo{author}{\bibfnamefont{J.}~\bibnamefont{Lekner}},
  \bibinfo{journal}{Physica~A} \textbf{\bibinfo{volume}{176}},
  \bibinfo{pages}{485} (\bibinfo{year}{1991}).

\bibitem[{\citenamefont{Schiessel and Pincus}(1998)}]{Schiessel-1998}
\bibinfo{author}{\bibfnamefont{H.}~\bibnamefont{Schiessel}} \bibnamefont{and}
  \bibinfo{author}{\bibfnamefont{P.}~\bibnamefont{Pincus}},
  \bibinfo{journal}{Macromolecules} \textbf{\bibinfo{volume}{31}},
  \bibinfo{pages}{7953} (\bibinfo{year}{1998}).

\bibitem[{\citenamefont{Doi and Edwards}(1986)}]{Doi-1986}
\bibinfo{author}{\bibfnamefont{M.}~\bibnamefont{Doi}} \bibnamefont{and}
  \bibinfo{author}{\bibfnamefont{S.~F.} \bibnamefont{Edwards}},
  \emph{\bibinfo{title}{The Theory of Polymer Dynamics}}
  (\bibinfo{publisher}{Oxford University Press}, \bibinfo{address}{Oxford},
  \bibinfo{year}{1986}).

\bibitem[{\citenamefont{Zifferer}(1998)}]{Zifferer-1998}
\bibinfo{author}{\bibfnamefont{G.}~\bibnamefont{Zifferer}},
  \bibinfo{journal}{J. Chem. Phys.} \textbf{\bibinfo{volume}{109}},
  \bibinfo{pages}{3691} (\bibinfo{year}{1998}).

\bibitem[{\citenamefont{Th\"{u}nemann}(1997)}]{Thuene2}
\bibinfo{author}{\bibfnamefont{A.}~\bibnamefont{Th\"{u}nemann}},
  \bibinfo{journal}{Langmuir} \textbf{\bibinfo{volume}{13}},
  \bibinfo{pages}{6040} (\bibinfo{year}{1997}).

\bibitem[{\citenamefont{Groot}(2003)}]{Groot-2003}
\bibinfo{author}{\bibfnamefont{R.~D.} \bibnamefont{Groot}},
  \bibinfo{journal}{J. Chem. Phys.} \textbf{\bibinfo{volume}{118}},
  \bibinfo{pages}{11265} (\bibinfo{year}{2003}).

\bibitem[{\citenamefont{Messina
  et~al.}(2002{\natexlab{b}})\citenamefont{Messina, Holm, and
  Kremer}}]{Messina-2002b}
\bibinfo{author}{\bibfnamefont{R.}~\bibnamefont{Messina}},
  \bibinfo{author}{\bibfnamefont{C.}~\bibnamefont{Holm}}, \bibnamefont{and}
  \bibinfo{author}{\bibfnamefont{K.}~\bibnamefont{Kremer}},
  \bibinfo{journal}{J. Chem. Phys.} \textbf{\bibinfo{volume}{117}},
  \bibinfo{pages}{2947} (\bibinfo{year}{2002}{\natexlab{b}}).

\bibitem[{\citenamefont{Morris et~al.}(1996)\citenamefont{Morris, Deaven, , and
  Ho}}]{Morris-1996}
\bibinfo{author}{\bibfnamefont{J.~R.} \bibnamefont{Morris}},
  \bibinfo{author}{\bibfnamefont{D.~M.} \bibnamefont{Deaven}}, ,
  \bibnamefont{and} \bibinfo{author}{\bibfnamefont{K.~M.} \bibnamefont{Ho}},
  \bibinfo{journal}{Phys. Rev.~B} \textbf{\bibinfo{volume}{53}},
  \bibinfo{pages}{R1740} (\bibinfo{year}{1996}).

\end{thebibliography}
\end{document}